\documentclass[journal=jacsat,manuscript=article]{achemso}

\usepackage{graphicx}
\usepackage{amsmath}
\usepackage{bm}
\usepackage{amssymb}
\usepackage{units}
\usepackage{color}
\usepackage{setspace}
\usepackage{braket}

\DeclareMathOperator{\erf}{erf}

\author{R. Krause}
\email{razvan.krause@ur.de}
\affiliation{University of Regensburg, Institute for Experimental and Applied Physics, Regensburg, Germany}
\alsoaffiliation{Max Planck Institute for the Structure and Dynamics of Matter, Center for Free Electron Laser Science, Hamburg, Germany}

\author{S. Aeschlimann}
\affiliation{University of Regensburg, Institute for Experimental and Applied Physics, Regensburg, Germany}
\alsoaffiliation{Max Planck Institute for the Structure and Dynamics of Matter, Center for Free Electron Laser Science, Hamburg, Germany}

\author{M. Ch\'avez-Cervantes}
\affiliation{Max Planck Institute for the Structure and Dynamics of Matter, Center for Free Electron Laser Science, Hamburg, Germany}

\author{R. Perea-Causin}
\affiliation{Department of Physics, Chalmers University of Technology, Gothenburg, Sweden}
\author{S. Brem}
\affiliation{Department of Physics, Philipps-Universit\"at Marburg, Renthof 7,
35032 Marburg, Germany}
\author{E. Malic}
\affiliation{Department of Physics, Philipps-Universit\"at Marburg, Renthof 7,
35032 Marburg, Germany}
\alsoaffiliation{Department of Physics, Chalmers University of Technology, Gothenburg, Sweden}

\author{S. Forti}
\affiliation{Center for Nanotechnology Innovation at NEST, Istituto Italiano di Tecnologia, Pisa, Italy}
\author{F. Fabbri}
\affiliation{Center for Nanotechnology Innovation at NEST, Istituto Italiano di Tecnologia, Pisa, Italy}
\alsoaffiliation{NEST, Istituto Nanoscienze, CNR and Scuola Normale Superiore, Pisa, Italy}
\alsoaffiliation{Graphene Labs, Istituto Italiano di Tecnologia, Genova, Italy}
\author{C. Coletti}
\affiliation{Center for Nanotechnology Innovation at NEST, Istituto Italiano di Tecnologia, Pisa, Italy}
\alsoaffiliation{Graphene Labs, Istituto Italiano di Tecnologia, Genova, Italy}

\author{I. Gierz}
\email{isabella.gierz@ur.de}
\affiliation{University of Regensburg, Institute for Experimental and Applied Physics, Regensburg, Germany}

\title{Microscopic understanding of ultrafast charge transfer in van-der-Waals heterostructures}

\abbreviations{tr-ARPES}

\begin{document}






\pagebreak

\begin{abstract}
Van-der-Waals heterostructures show many intriguing phenomena including ultrafast charge separation following strong excitonic absorption in the visible spectral range. However, despite the enormous potential for future applications in the field of optoelectronics, the underlying microscopic mechanism remains controversial. Here we use time- and angle-resolved photoemission spectroscopy combined with microscopic many-particle theory to reveal the relevant microscopic charge transfer channels in epitaxial WS$_2$/graphene heterostructures. We find that the timescale for efficient ultrafast charge separation in the material is determined by direct tunneling at those points in the Brillouin zone where WS$_2$ and graphene bands cross, while the lifetime of the charge separated transient state is set by defect-assisted tunneling through localized sulphur vacanices. The subtle interplay of intrinsic and defect-related charge transfer channels revealed in the present work can be exploited for the design of highly efficient light harvesting and detecting devices.
\end{abstract}

\pagebreak



Stacking different two-dimensional materials in a lego-like manner enables the formation of novel ultimately thin heterostructures with tailored electronic properties exploiting screening and proximity-induced effects \cite{Geim_Nature2013, Novoselov_Science2016, Merkl_NatMater2019}. These van-der-Waals (vdW) heterostructures exhibit many intriguing phenomena including ultrafast charge separation following optical excitation \cite{Jin_NatNanotechnol2018} and bear great promise for future applications in the field of optoelectronics \cite{Liu_NatRevMater2106}. Among the huge variety of existing vdW heterostructures those that combine monolayer graphene with a monolayer of one of the semiconducting transition metal dichalcogenides (TMDs) are of particular interest \cite{Lorchat_NatNanotechnol2020}. These heterostructures exhibit type I band alignment where both the minimum of the conduction band (CBM) and the maximum of the valence band (VBM) are located in the graphene layer. Strong excitonic absorption in the TMD monolayer is then followed by ultrafast charge transfer into the graphene layer \cite{He_NatCommuun2014, Huo_JMaterChemC2015, Massicotte_NatNanotechnol2016, Hill_PRB2017, Yuan_SciAdv2018, He_OptExpr2017, Song_Optik2018}. Previous studies on WS$_2$/graphene heterostructures have shown that hole transfer is considerably faster than electron transfer, resulting in a charge-separated transient state \cite{Aeschlimann_SciAdv2020,Fu_arXiv2020}. In addition, broken inversion symmetry together with strong spin-orbit coupling of the TMD monolayer allows for spin-selective optical excitation \cite{Yao_PRB2008, Xiao_PRL2012, Cao_NatCommun2012, Zeng_NatNanotechnol2012, Mak_NatNanotechnol2012} such that ultrafast charge transfer might be accompanied by spin transfer, allowing for optical spin injection into graphene \cite{Gmitra_PRB2015, Luo_NanoLett2017, Avsar_ACSNano2017}. Despite the enormous potential for future applications in the field of optoelectronics and optospintronics, the microscopic mechanism underlying the ultrafast charge transfer processes in TMD/graphene and similar heterostructures remains poorly understood.

In this work we tackle this problem with a combination of time- and angle-resolved photoemission spectroscopy (tr-ARPES) and microscopic many-particle theory \cite{Malic_Book2013, Brem_NanoLett2020, Brem_Nanoscale2020} where the measured asymmetry of the population dynamics above and below the Fermi level as well as transient band shifts serve as smoking gun evidence for ultrafast charge separation \cite{Aeschlimann_SciAdv2020}. We find that both electron and hole transfer become faster with increasing pump fluence for resonant excitation of the A-exciton in WS$_2$ of our epitaxial WS$_2$/graphene hetrostructures \cite{Forti_Nanoscale2017, Fabbri_JPhysChemC2020}. We explain this with the help of a tunneling model where hot carriers transfer from WS$_2$ to graphene at those points in momentum space where the respective bands cross and where charge transfer becomes possible without energy or momentum transfer. These crossing points are separated from the VBM and CBM of WS$_2$ by an energy barrier. The higher the non-equilibrium carrier temperature, the easier it becomes for the carriers to overcome these barriers and to tunnel into the graphene layer. Surprisingly, we also find that the lifetime of the charge separated state increases with increasing fluence. This trend is not captured by our microscopic simulation indicating the importance of defects. We propose that in-gap states related to S vacancies \cite{Wei_AIPAdv2012, Carozo_SciAdv2017, Schuler_PRL2019} efficiently trap photoexcited electrons in the WS$_2$ layer \cite{Fu_arXiv2020} and thereby set the lifetime for the charge separated state. The microscopic insights gained in the present study will guide the design of future optoelectronic and optospintronic devices where defects and band alignment will serve as important control parameters.


WS$_2$/graphene heterostructures were grown on H-terminated SiC(0001) as described in Refs. \cite{Forti_Nanoscale2017} and \cite{Fabbri_JPhysChemC2020}. 2\,eV pump and 26\,eV probe pulses for the tr-ARPES experiments were generated by frequency doubling the signal output of an optical parametric amplifier and by high harmonics generation in Argon, respectively. The tunneling matrix elements for electrons and holes were computed using microscopic many-particle theory \cite{Malic_Book2013, Brem_NanoLett2020, Brem_Nanoscale2020}. Further details about the various methods employed in the present study are given in the Supporting Information. 

\begin{figure*}
	\center
		\includegraphics[width = 1\columnwidth]{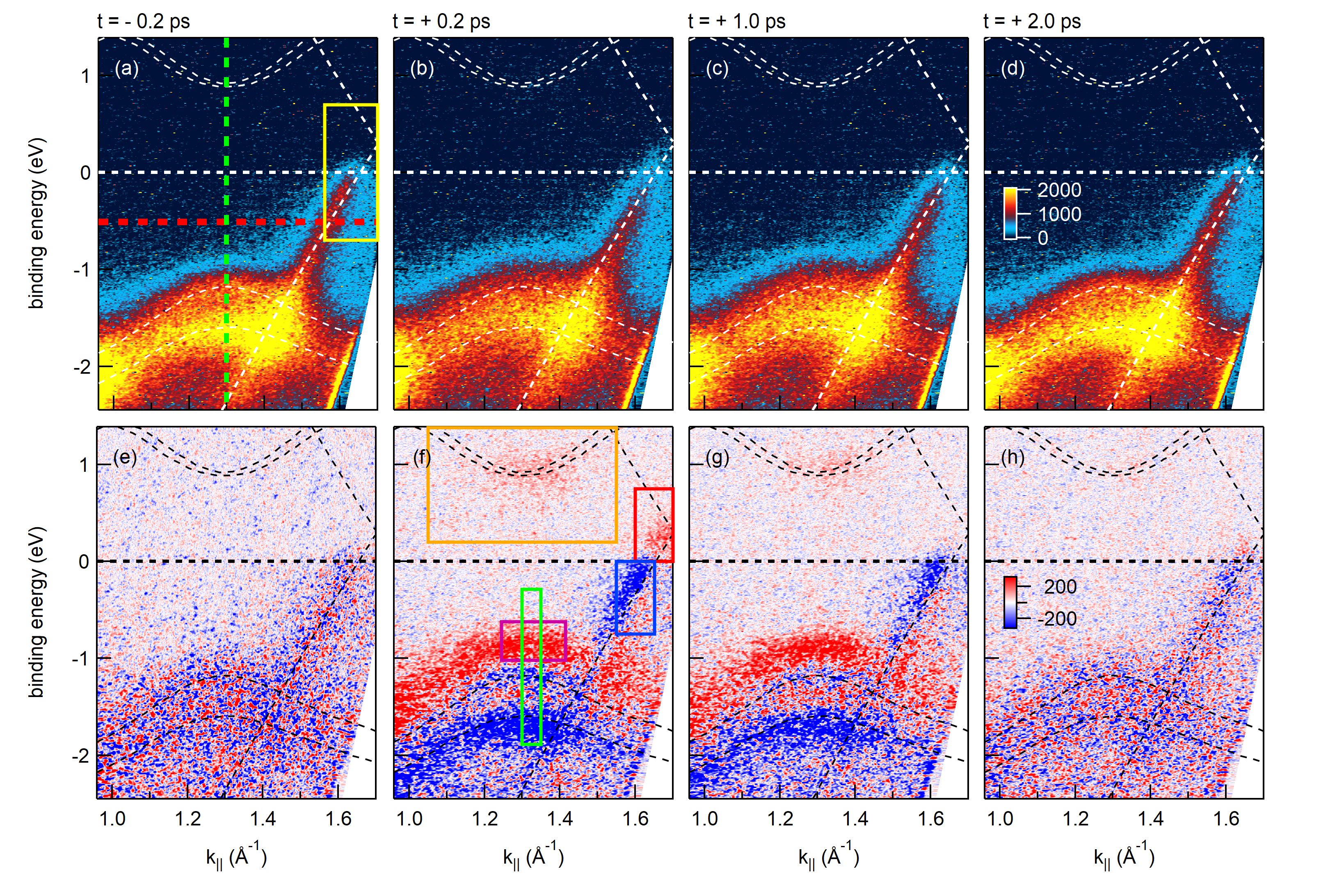}
  \caption{\textbf{tr-ARPES snapshots of WS$_2$/graphene heterostructure before and after optical excitation:} (a) Photocurrent measured along the $\Gamma$K-direction at negative pump-probe delay. Colored dashed lines indicate the position of the line profiles used to determine the transient band positions in Fig.\,\ref{Figure2}c and e. The yellow box indicates the region where the Fermi-Dirac fits for Fig.\,\ref{Figure4}c were performed. (b)-(d) Photocurrent for different pump-probe time delays after photoexcitation at $\hbar\omega_{\text{pump}}=2$\,eV with a pump fluence of 2.85\,mJ/cm$^2$. (e)-(h) Pump-induced changes of the photocurrent for different pump-probe delays. Red and blue indicate gain and loss of photoelectrons, respectively, with respect to the photocurrent measured at negative pump-probe delay. The colored boxes in (f) indicate the area of integration for the pump-probe traces shown in Fig.\,\ref{Figure2}a and b. Thin white (a-d) and black (e-h) dashed lines represent the calculated band structures of graphene \cite{Wallace_PR1947} and WS$_2$ \cite{Zeng_SciRep2013}. A rigid band shift of $-0.81$\,eV\ and $-1.19$\,eV was applied to the WS$_2$ CB and VB, respectively. The graphene Dirac cone was shifted by $+0.3$\,eV to account for the observed hole doping \cite{Riedl_PRL2009}. Thick white (a-d) and black (e-h) dashed lines mark the position of the Fermi level.}
  \label{Figure1}
\end{figure*}

In Fig.\,\ref{Figure1} we show tr-ARPES snapshots measured along the $\Gamma$K-direction of the hexagonal Brillouin zone close to the graphene and WS$_2$ K-points for different pump-probe delays after photoexcitation at $\hbar\omega_{\text{pump}}=2$\,eV with a pump fluence of 2.85\,mJ/cm$^2$ (Fig.\,\ref{Figure1}a-d) together with the pump-induced changes of the photocurrent (Fig.\,\ref{Figure1}e-h). Within our experimental resolution (170\,meV in this particular data set) the unperturbed band structure measured at negative pump-probe delay (Fig.\,\ref{Figure1}a) is well described by the sum of the calculated band structures of the individual layers \cite{Wallace_PR1947,Zeng_SciRep2013} (thin white dashed lines in Fig.\,\ref{Figure1}a) after applying rigid band shifts to account for doping and to reproduce the experimentally observed WS$_2$ band gap. At the peak of the pump-probe signal ($t=0.2$\,ps in Fig.\,\ref{Figure1}b and f) we observe a gain of photoelectrons at the bottom of the WS$_2$ conduction band (CB), a strong gain-and-loss signal for the WS$_2$ valence band (VB), and a strong gain-and-loss signal for the graphene Dirac cone. We interpret these features in terms of electron-hole pair generation followed by ultrafast charge separation as discussed in detail below \cite{Aeschlimann_SciAdv2020}. 

\begin{figure*}
	\center
		\includegraphics[width = 1\columnwidth]{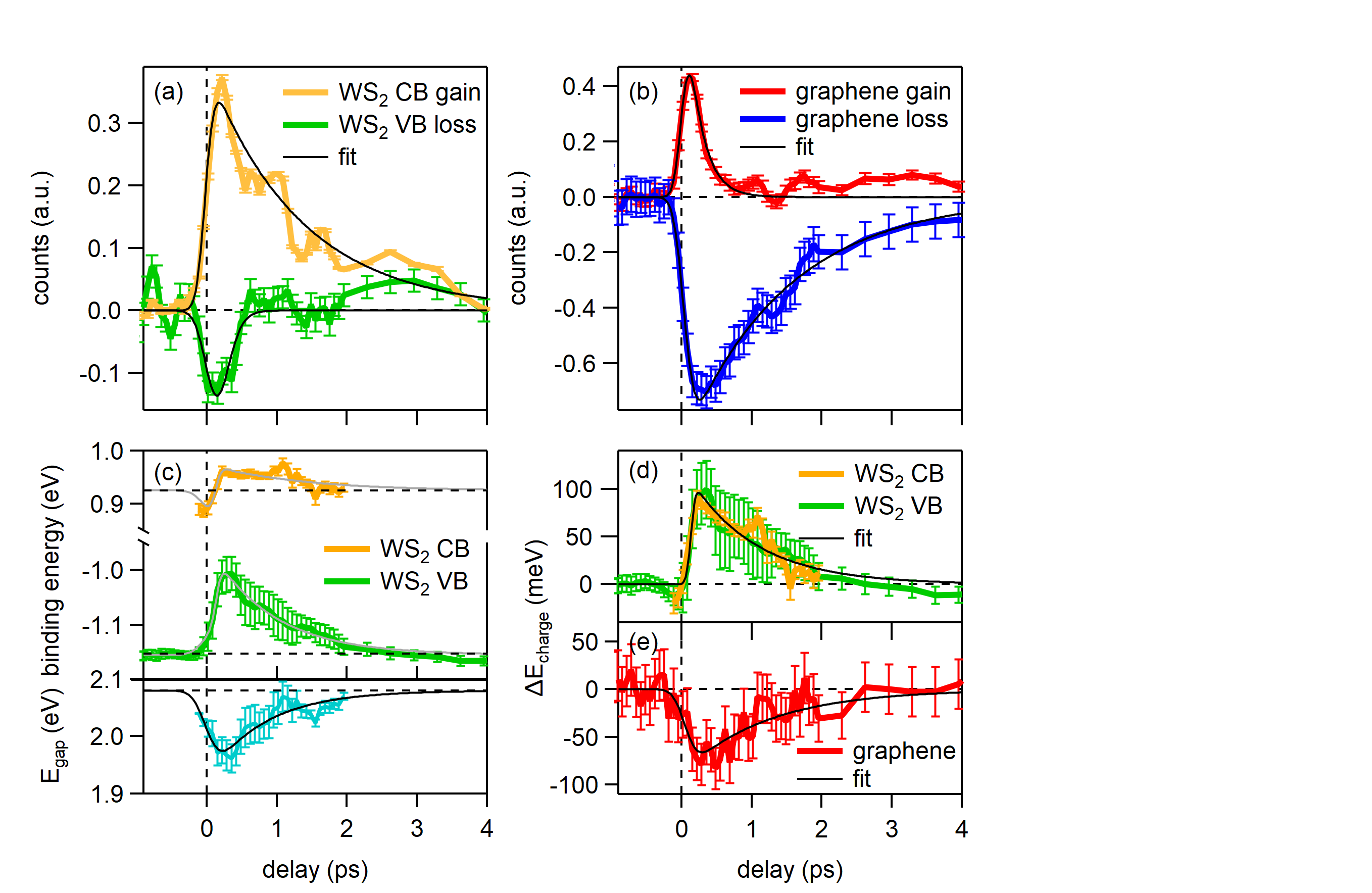}
  \caption{\textbf{smoking gun evidence for ultrafast charge separation:} (a) Gain in the WS$_2$ CB (orange) and loss in the WS$_2$ VB (green) as a function of pump-probe delay. (b) Gain above the Fermi level (red) and loss below the Fermi level (blue) inside the graphene Dirac cone as a function of pump-probe delay. (c) Position of the WS$_2$ CB (orange) and the upper WS$_2$ VB (green) as a function of pump-probe delay. The transient band gap of WS$_2$ (turquois) was obtained by subtracting the transient position of the VB from the transient position of the CB. (d) Charging-induced shift of WS$_2$ valence and conduction band obtained by adding (subtracting) $|\Delta E_{\text{gap}}|/2$ to (from) the VB (CB) position in panel a. (e) Transient shift of the graphene Dirac cone. Black lines are single exponential fits to the data. Gray lines in panel c were calculated from the exponential fit for $E_{\text{gap}}$ in panel c and the exponential fit in panel d. All data was obtained with a pump fluence of 2.85\,mJ/cm$^2$.}
  \label{Figure2}
\end{figure*}

For further analysis we present momentum-resolved population dynamics (Fig.\,\ref{Figure2}a and b) as well as transient peak positions of the individual electronic bands (Fig.\,\ref{Figure2}c and d) that serve as smoking gun evidence for ultrafast charge separation in the system. The data points in Fig. \ref{Figure2}a and b were obtained by integrating the photocurrent over the areas indicated by the colored boxes in Fig. \ref{Figure1}f. They show the population dynamics of the WS$_2$ valence and conduction band (Fig.\,\ref{Figure2}a) and the population dynamics of the graphene Dirac cone (Fig.\,\ref{Figure2}b) as a function of pump-probe delay. Thin black lines are single exponential fits to the data (see Supporting Information section II). We find that gain and loss traces for both WS$_2$ and graphene are asymmetric. The WS$_2$ valence and conduction band show a short-lived loss ($\tau=100\pm40$\,fs) and long-lived gain ($\tau=1.3\pm0.1$\,ps), respectively. The situation is reversed in the graphene layer where the gain above the Fermi level is found to be short-lived ($\tau=210\pm20$\,fs) and the loss below the Fermi level is found to be long-lived ($\tau=1.50\pm0.06$\,ps). This behavior is not observed in individual graphene and WS$_2$ layers and indicates ultrafast charge separation \cite{Aeschlimann_SciAdv2020}. In detail, the short lifetime of the loss in the WS$_2$ VB and the gain above the Fermi level in the graphene Dirac cone indicate that the photoexcited holes in the WS$_2$ layer are rapidly (on a timescale comparable to the temporal resolution of 200\,fs) refilled by electrons from graphene. The photoexcited electrons, on the other hand, are found to remain in the WS$_2$ CB for $\tau=1.3\pm0.1$\,ps. 

The resulting charge separated transient state leaves the WS$_2$ layer negatively charged and the graphene layer positively charged. This is expected to decrease the binding energy of the WS$_2$ states and increase the binding energy of the graphene states \cite{Aeschlimann_SciAdv2020}. In Fig.\,\ref{Figure2}c we plot the transient peak positions of the upper WS$_2$ VB and the WS$_2$ CB that were obtained by fitting energy distribution curves (EDCs) through the K-point of WS$_2$ (dashed green line in Fig.\,\ref{Figure1}a) with a Gaussian (see Supporting Information section II). The energy difference between the WS$_2$ CB and VB directly yields the transient band gap $E_{\text{gap}}$ shown in the lower part of Fig.\,\ref{Figure2}c. From an exponential fit to the data (see Supporting Information section II) we deduce an equilibrium gap size of 2.08\,eV and a lifetime of $\tau=0.8\pm 0.4$\,ps for the transient band gap renormalization $\Delta E_{\text{gap}}$. The renormalization is a consequence of increased screening in the presence of photoexcited carriers and is commonly observed in photodoped TMD monolayers \cite{Ugeda_NatMater2014, Chernikov_NatPhoton2015, Ulstrup_ACSNano2016, Liu_PRL2019}. Assuming that band gap renormalization shifts the WS$_2$ VB up and the WS$_2$ CB down by the same amount $|\Delta E_{\text{gap}}|/2$ we can subtract its contribution from the transient peak positions in Fig.\,\ref{Figure2}c yielding the data in Fig.\,\ref{Figure2}d where both the WS$_2$ valence and conduction band are found to shift up by $\sim$100\,meV with a lifetime of $\tau=1.03\pm0.07$\,ps. This shift is attributed to the transient negative charging of the WS$_2$ layer. The transient position of graphene's Dirac cone in Fig.\,\ref{Figure2}e was extracted from Lorentzian fits of momentum distribution curves (MDCs) at $E=-0.5$\,eV extracted along the dashed red line in Fig.\,\ref{Figure1}a. The resulting momentum shift was converted into an energy shift by multiplying with the slope of the band $\nu_F=7$\,eV\AA \cite{CastroNeto_RevModPhys2009}. The observed increase in binding energy with an exponential lifetime of $1.08\pm0.06$\,ps is a direct consequence of the additional positive charge on the graphene layer due to ultrafast hole transfer.

\begin{figure*}
	\center
		\includegraphics[width = 1\columnwidth]{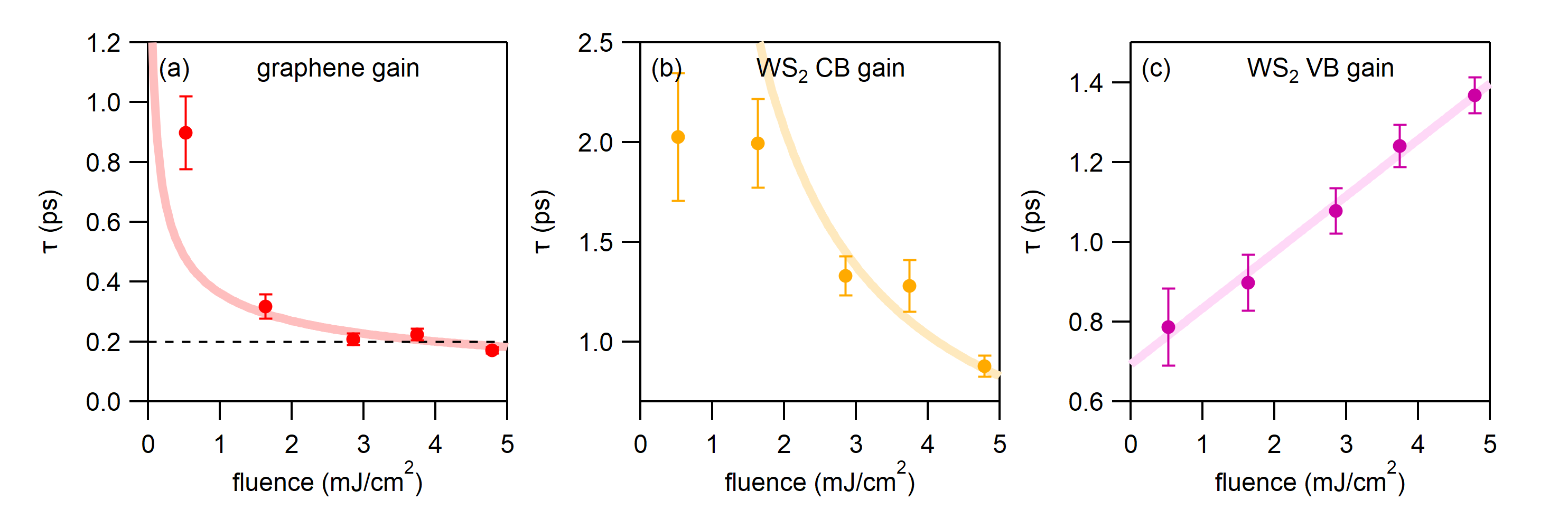}
  \caption{\textbf{fluence dependence of ultrafast charge separation and recombination:} (a) Lifetime of graphene gain from Fig.\,\ref{Figure2}b as a function of fluence. The black dashed line represents our temporal resolution of 200\,fs. (b) Lifetime of photoexcited electrons in the WS$_2$ CB from Fig.\,\ref{Figure2}a as a function of fluence. (c) Lifetime of gain above the equilibrium position of the upper WS$_2$ VB obtained by integrating the photocurrent over the area marked by the pink box in Fig.\,\ref{Figure1}f and by fitting the resulting population dynamics with a single exponential decay. Thick lines in a and b were obtained by inverting and rescaling the guides to the eye from Fig.\,\ref{Figure4}e to match the data points. The thick line in c is a guide to the eye.}
  \label{Figure3}
\end{figure*}

In order to gain access to the microscopic mechanism underlying the observed ultrafast charge transfer phenomena we now investigate the pump fluence dependence of the electron and hole transfer rates as well as the lifetime of the charge separated state. As discussed above, the short-lived loss in the WS$_2$ VB (Fig.\,\ref{Figure2}a) is directly linked to the short-lived gain above the Fermi level in the Dirac cone of graphene (Fig.\,\ref{Figure2}b). The signal-to-noise ratio, however, is much better in the latter case, which is why we focus on the graphene gain rather than the WS$_2$ VB loss to analyze the pump fluence dependence of the ultrafast hole transfer. The result is shown in Fig.\,\ref{Figure3}a. We find that the hole transfer time of $\sim1$\,ps observed at the lowest pump fluence rapidly drops to a value short compared to our temporal resolution of 200\,fs with increasing fluence. Figure\,\ref{Figure3}b shows the pump fluence dependence of the lifetime of the photoexcited electrons in the conduction band of WS$_2$ from Fig.\,\ref{Figure2}a. This value is found to decrease linearly from $\sim2$\,ps at the lowest fluence to below $1$\,ps at the highest fluence. In Fig.\,\ref{Figure3}c we plot the pump fluence dependence of the gain above the equilibrium position of the upper WS$_2$ VB obtained by integrating the photocurrent over the area indicated by the pink box in Fig. \ref{Figure1}f. We confirmed that this quantity shows the same fluence dependence as the charging-induced band shifts in Fig.\,\ref{Figure2}b (see Supporting Information section II) that are directly linked to the charge separated state albeit with an improved signal-to-noise ratio. From Fig.\,\ref{Figure3}c we find that the lifetime of the charge separated state linearly increases with increasing fluence from $\sim0.8$\,ps at the lowest fluence to $\sim1.4$\,ps at the highest fluence. In summary, Fig.\,\ref{Figure3} shows that the transfer rate of the holes increases with increasing fluence, the lifetime of the photoexcited electrons in the CB of WS$_2$ decreases with increasing fluence, and the lifetime of the charge separated state increases with increasing fluence. 

\begin{figure*}
	\center
		\includegraphics[width = 1\columnwidth]{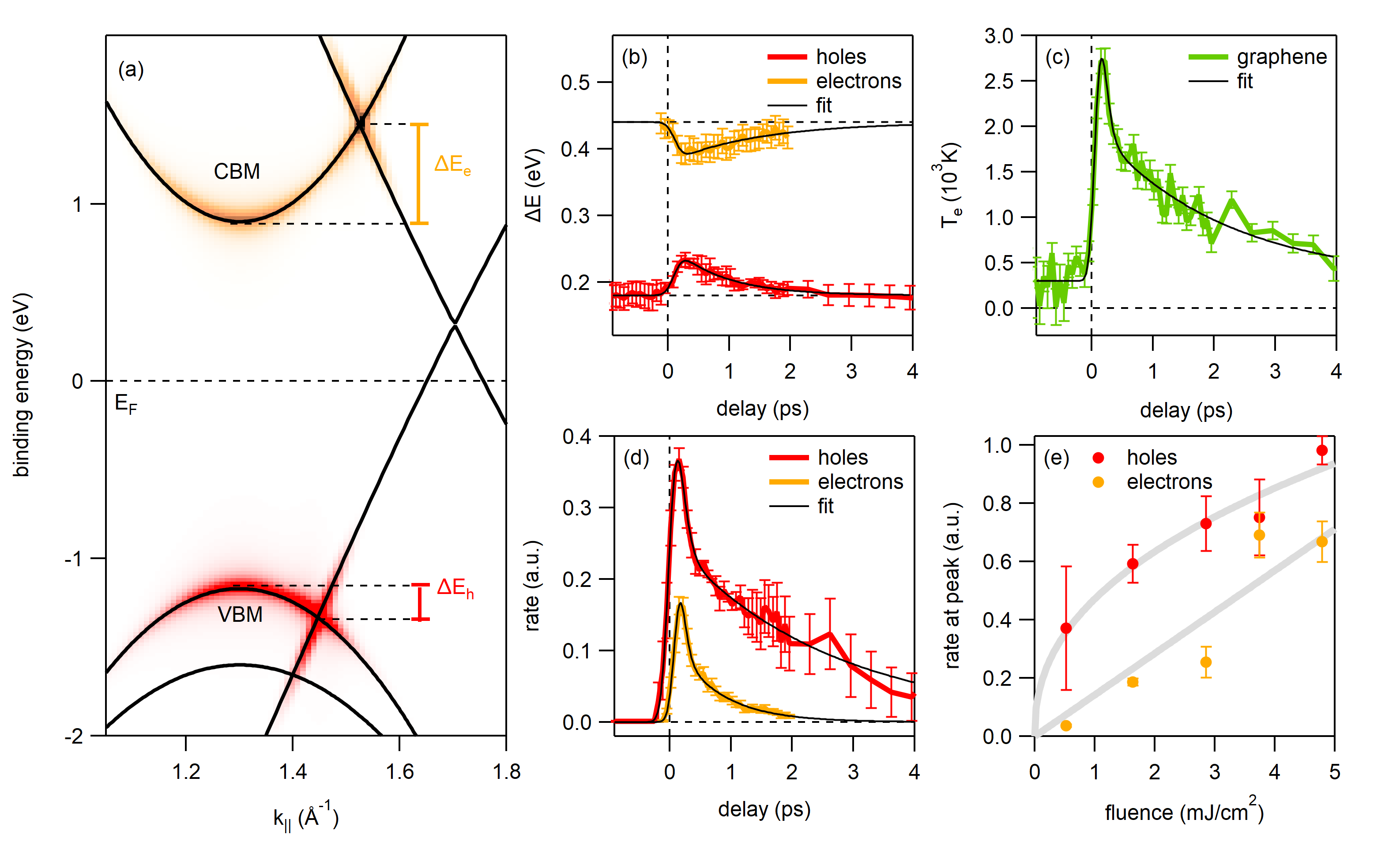}
  \caption{\textbf{direct tunneling model:} (a) Sketch of the band structure with tunneling barrier for holes ($\Delta E_h$, red) and electrons ($\Delta E_e$, orange). (b) Temporal evolution of the two barrier heights deduced from Fig.\,\ref{Figure2} together with single-exponential fit. (c) Temporal evolution of the electronic temperature inside the Dirac cone together with double-exponential fit. (d) Temporal evolution of the electron and hole transfer rates deduced from panels b and c together with double-exponential fit. (e) Pump fluence dependence of the peak transfer rates for electrons and holes. Thick gray lines are guides to the eye.}
  \label{Figure4}
\end{figure*}

A detailed comparison with related experimental \cite{Ulstrup_ACSNano2016, Song_Optik2018} and theoretical \cite{Long_NanoLett2016, Wang_NatCommun2016, Zheng_NanoLett2017} work from literature is given in the Supporting Information section III. At this point it suffices to say that none of the proposed models can account for the experimentally observed fluence dependence shown in Fig.\,\ref{Figure3}. Therefore, we evoke a new scenario where carriers can tunnel from one layer to the other at the points in momentum space where the bands of the individual layers cross and where charge transfer can occur without energy or momentum transfer. In this scenario the transfer rate is proportional to $e^{\Delta E/k_BT}$, where the energy barrier $\Delta E$ is given by the distance between the WS$_2$ VBM or CBM and the closest crossing point (see Fig.\,\ref{Figure4}a) and where $k_BT$ is the thermal energy of the carriers. The temporal evolution of the energy barriers for electron and hole transfer was deduced from the transient band shifts in Fig.\,\ref{Figure2} and is shown in Fig.\,\ref{Figure4}b. The temperature of the electron-hole pairs in WS$_2$ is difficult to determine experimentally due to the short lifetime of the holes and the limited signal-to-noise ratio. The electronic temperature of the Dirac carriers inside the graphene layer (see Fig.\,\ref{Figure4}c) on the other hand can be easily extracted from Fermi-Dirac fits of the transient carrier distribution inside the Dirac cone as described in the Supplementary Information in section II. Assuming that the electronic temperature for the carriers inside the Dirac cone is similar to the one of the carriers in the WS$_2$ layer, we can determine the temporal evolution of the transfer rates shown in Fig.\,\ref{Figure4}d. The peak rate is found to be faster for holes than for electrons and to increase with increasing fluence for both electrons and holes (see Fig.\,\ref{Figure4}e) in good qualitative agreement with the data in Fig.\,\ref{Figure3}. 

\begin{figure}
	\center
		\includegraphics[width = 1\columnwidth]{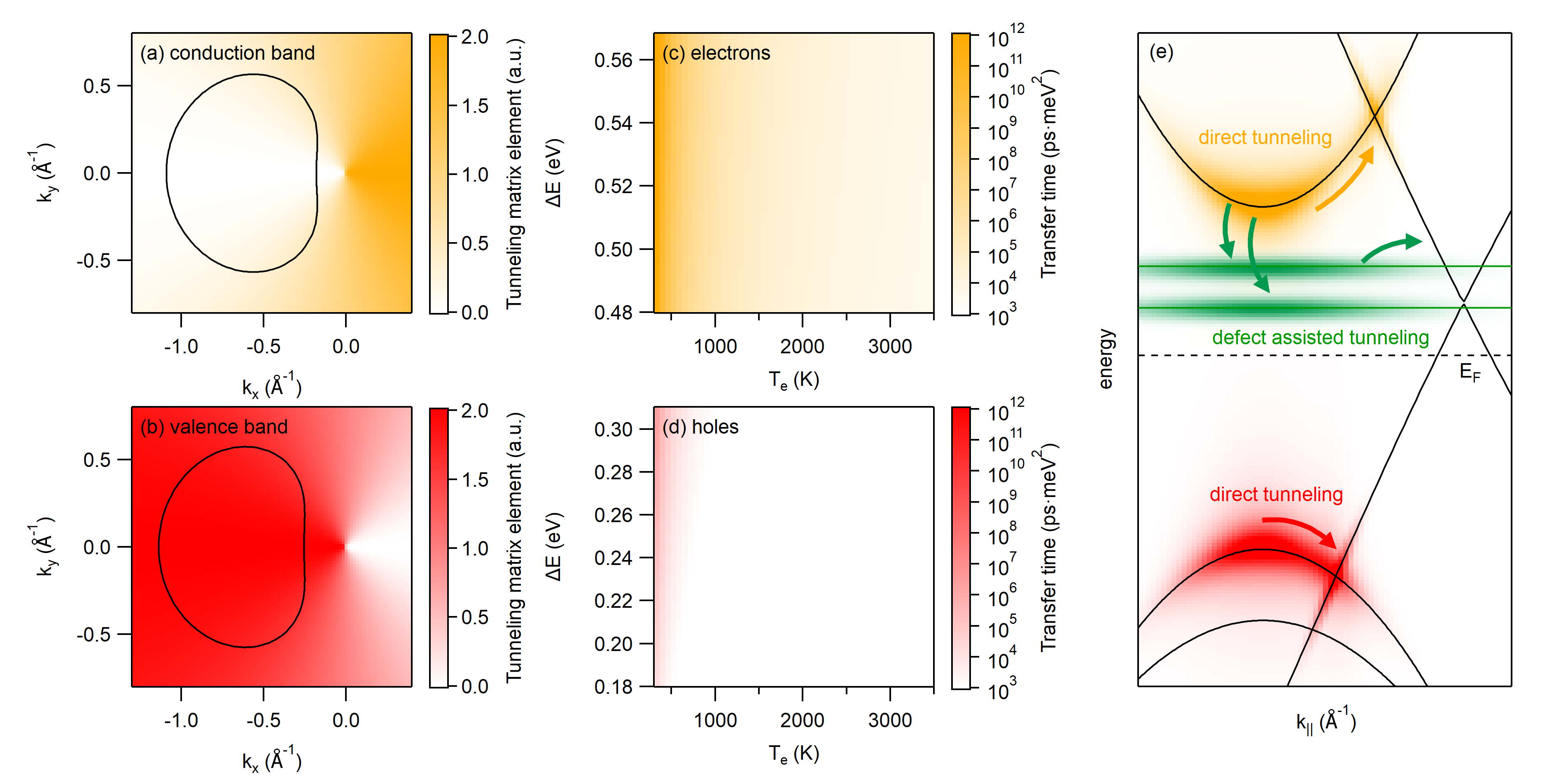}
  \caption{\textbf{theory and final microscopic model:} Momentum dependence of the tunneling matrix element calculated from tight-binding for CB (a) and VB (b). $k_x=k_y=0$ corresponds to the K-point of graphene. Calculated transfer time for electrons (c) and holes (d). (e) Final microscopic model including direct (yellow and red arrows) as well as defect-assisted tunneling (green arrows).}
  \label{Figure5}
\end{figure}

To confirm the above interpretation, we calculated the transfer times for electrons and holes using a microscopic model based on the density matrix formalism \cite{Malic_Book2013, Ovesen_CommunPhys2019, Brem_NanoLett2020, Brem_Nanoscale2020}, which is further described in the Supporting Information in section IV. The momentum dependence of the tunneling matrix element evaluated on a tight-binding level for CB and VB is shown in Fig.\,\ref{Figure5}a and b, respectively, together with black lines that indicate the position in momentum space where the crossing between WS$_2$ CB and VB, respectively, and the Dirac cone occurs. The tunneling matrix elements exhibit a strong momentum dependence that can be traced back to the pseudospin of the Dirac carriers in the graphene layer. More importantly, the tunneling matrix element in the region of interest is found to be much bigger for the VB than for the CB. In addition to the different energy barriers found in Fig.\,\ref{Figure3}b this accounts for the experimentally observed asymmetry between electron and hole transfer. We also calculated the transfer times for holes and electrons as a function of the height of the respective energy barrier and the carrier temperature in Fig.\,\ref{Figure5}c and d. In good qualitative agreement with our experiment, the calculated lifetimes are found to decrease with increasing carrier temperature (i.\,e. with increasing pump fluence), while the change in barrier height ($<100$\,meV in the experiment) is found to have a minor influence on the fluence dependence of the lifetimes. 

While this model nicely explains the observed fluence dependence of the carrier lifetimes in Fig.\,\ref{Figure3}a and b, it fails to reproduce the fact that the lifetime of the charge separated transient state is found to increase with increasing fluence (Fig.\,\ref{Figure3}c). This indicates that, in addition to the direct tunneling scenario proposed above, there are additional charge transfer channels that need to be considered for a full microscopic understanding of the ultrafast charge transfer processes in our WS$_2$/graphene heterostructure. A likely candidate are defect-related charge transfer channels involving S vacancies \cite{Carozo_SciAdv2017, Schuler_PRL2019}. It has been shown using scanning tunneling spectroscopy combined with GW calculations \cite{Schuler_PRL2019} that S vacancies give rise to two spin-orbit split states inside the band gap of WS$_2$. These states were proposed to efficiently trap photoexcited electrons inside the WS$_2$ layer and thereby enhance the lifetime of the charge separated state in commercial WS$_2$/ graphene heterostructures \cite{Fu_arXiv2020}. 

In Fig.\,\ref{Figure5}e we summarize our understanding of the relevant microscopic scattering channels that mediate the ultrafast charge transfer in our WS$_2$/graphene heterostructure. Direct tunneling (yellow and red arrows) sets the timescale for charge separation, while defect-assisted tunneling (blue arrows) sets the timescale for charge recombination. This model resolves two important controversies in literature. First, it remains unresolved whether charge transfer occurs at K where the interlayer coupling is relatively weak \cite{Wang_NatCommun2016, Zhang_AdvSci2017, Long_NanoLett2016, Li_ChemMater2017} or at other valleys ($\Gamma$ and $\Sigma$) where interlayer coupling is stronger \cite{Wang_PRB2017, Zheng_NanoLett2017}. Our findings suggest that ultrafast charge transfer in the investigated epitaxial WS$_2$/graphene heterostructure occurs at the band intersections close to the K-point. The second unresolved issue is related to momentum conservation during charge transfer \cite{Zhu_JACS2015, Wang_PRB2017}. We propose that ultrafast charge separation occurs via direct tunneling with zero momentum transfer. Charge carrier recombination is dominated by defect-assisted tunneling where the strong localization of the defect states in real space leads to a delocalization over a big part of the Brillouin zone which again enables tunneling at zero momentum transfer.


In summary we combined tr-ARPES with microscopic many-particle theory to unravel the microscopic mechanism of ultrafast charge separation and recombination in epitaxial WS$_2$/graphene heterostructures. We find a subtle interplay between direct tunneling at the band intersections close to the K-point that sets the timescale for charge separation and defect-assisted tunneling via localized S vacancies that sets the timescale for electron-hole recombination. Our findings will guide the development of improved optoelectronic devices where defect and band structure engineering will serve as important control parameters.

\begin{acknowledgement}
We thank S. Latini, L. Xian, A. Rubio, and S. Refaely-Abramson for many fruitful discussions. This  work  was  supported  by  the  Deutsche Forschungsgemeinschaft through SFB 925 and SFB 1277, by the European Union’s Horizon 2020 research and innovation program under grant agreement no. 785219 and no. 881603, and by the Swedish Research Council (VR, project number 2018-00734). The computations were enabled by resources provided by the Swedish National Infrastructure for Computing (SNIC) at C3SE partially funded by the Swedish Research Council through grant agreement no. 2016-07213. R.P.C. acknowledges funding from the Excellence Initiative Nano (Chalmers) under the Excellence PhD programme.
\end{acknowledgement}

\begin{suppinfo}
The Supporting Information contains details about sample growth, tr-ARPES setup, data analysis, and simulations.
\end{suppinfo}


\clearpage
\pagebreak

\clearpage
\pagebreak

\section{Supporting Information}

\section{I Sample growth and characterization}  

\begin{figure}
	\center
		\includegraphics[width = 1\columnwidth]{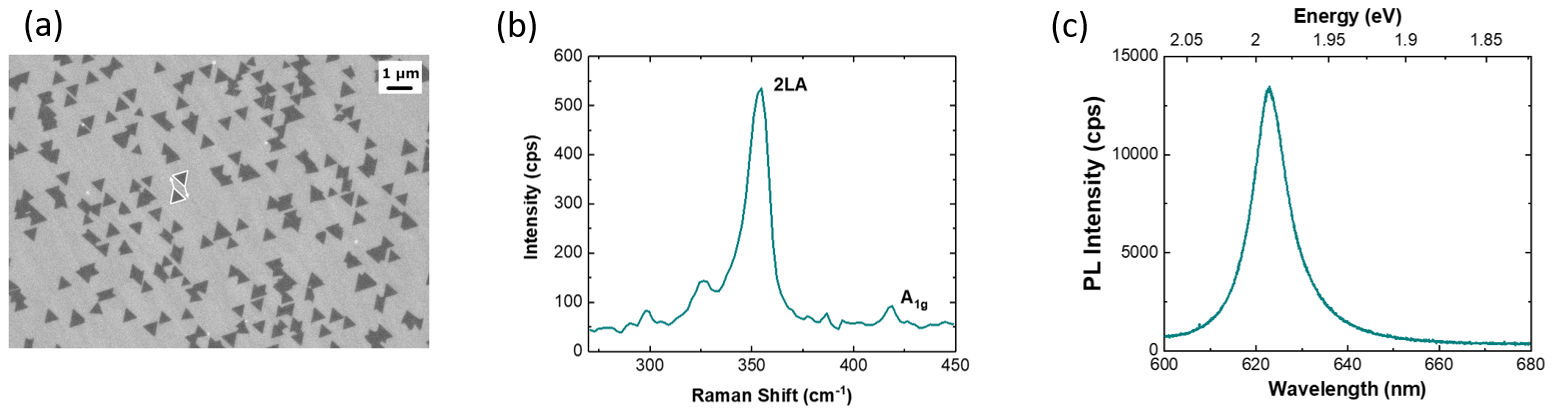}
  \caption{\textbf{Sample characterization:} (a) Secondary electron microscopy image. (b) Raman spectrum. (c) Photoluminescence spectrum.}
  \label{SFig1}
\end{figure}

\noindent Graphene samples were grown on N-doped 6H-SiC(0001) with a miscut below 0.5$^{\circ}$ bought from SiCrystal GmbH. The substrate was first cleaned and smoothed via etching in hydrogen atmosphere \cite{} and then graphitized by annealing in argon atmosphere \cite{EmtsevNatMater2009} until a carbon buffer layer with a $(6\sqrt{3}\times6\sqrt{3})$R30$^{\circ}$ surface reconstruction was formed. The carbon buffer layer was then decoupled from the substrate via hydrogen intercalation at 800$^{\circ}$C \cite{RiedlPRL2009} in order to obtain a quasi-free-standing graphene monolayer. The whole process was carried out in a commercial Black Magic™ reactor from Aixtron. The WS$_2$ growth was carried out in a standard hot-wall reactor by chemical vapor deposition (CVD) \cite{}. WO$_3$ and S powders with a weight ratio of 1:50 were used as precursors. The WO$_3$ and S powders were kept at 900$^{\circ}$C and 120$^{\circ}$C, respectively. The WO$_3$ powder was placed close to the substrate. Argon was used as carrier gas with a flow of 80\,sccm. The pressure in the reactor was kept at 1\,mbar. The synthesis process took 30\,min. The samples were characterized with secondary electron microscopy, Raman, and photoluminescence spectroscopy. The results are shown in Fig.\,\ref{SFig1}.

\section{II Time- and angle-resolved photoemission spectroscopy} 

\subsection{experimental setup}

\begin{figure}
	\center
		\includegraphics[width = 0.5\columnwidth]{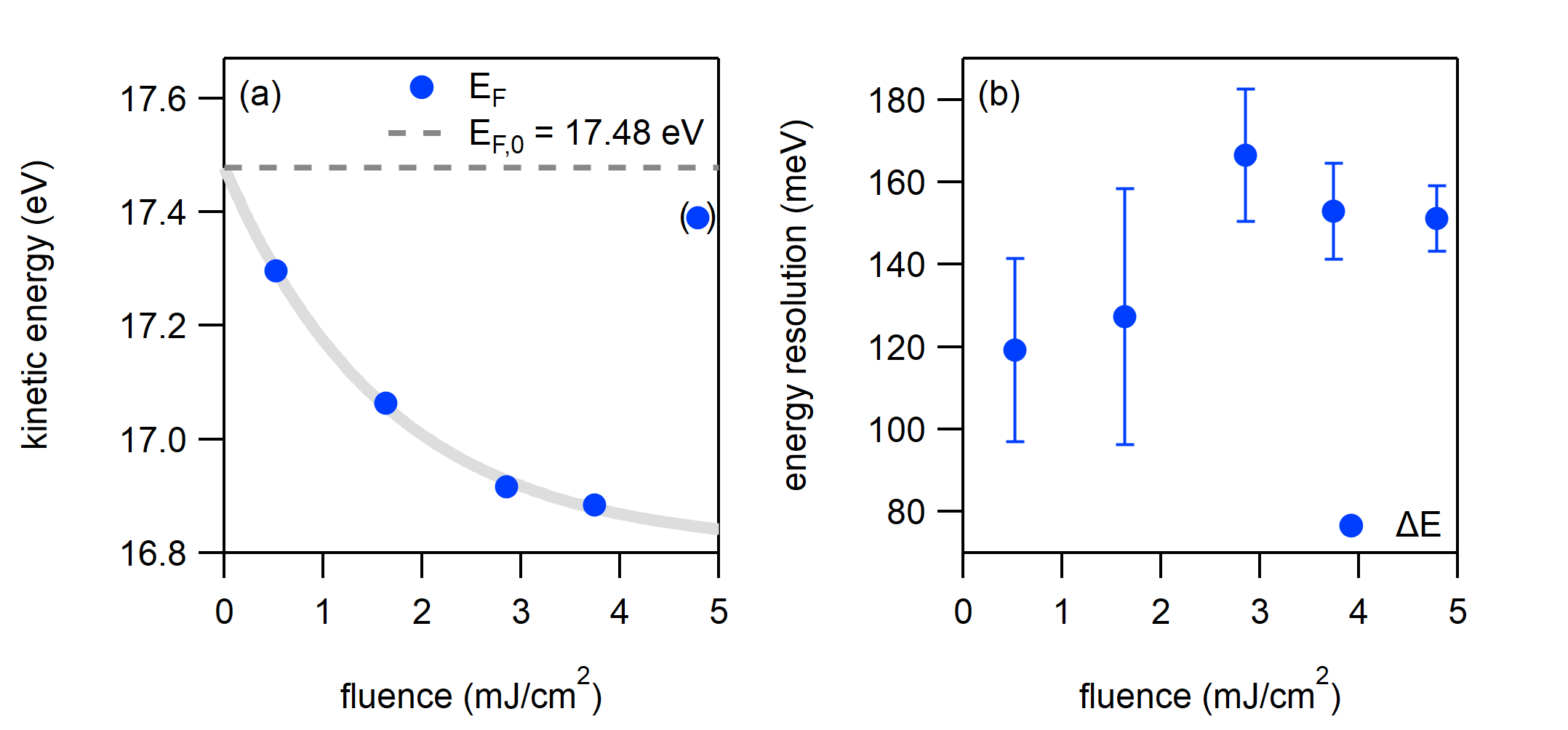}
  \caption{\textbf{Space charge effects:} (a) Position of the Fermi level at negative delays as a function of pump fluence. The gray line is used to extrapolate the position of the Fermi level to zero pump fluence. The data point in brackets was obtained on an area of the sample where the WS$_2$ coverage was particularly low. The error bars are smaller than the symbol size. (b) Energy resolution for different pump fluences.}
  \label{SFig2}
\end{figure}

\noindent We performed the tr-ARPES experiments in an 2\,eV-pump/XUV-tr-ARPES-probe setup based on a 1\,kHz Titanium:Sapphire amplifier (Coherent Legend Elite Duo). Extreme ultraviolet (XUV) pulses were obtained from high harmonics generation (HHG) in an argon gas jet. Probe pulses at 26\,eV photon energy with a nominal pulse duration of 100\,fs were selected with a grating monmochromator. Pump pulses were obtained by frequency doubling of the signal output of an optical parametric amplifier (HE-TOPAS from Light Conversion). The pump fluence range  employed in the present study was limited by noise on the lower side and pump-induced space charge on the upper side. ARPES snapshots were measured with a hemispherical analyzer (SPECS Phoibos\,100). The photocurrent is detected on a two-dimensional detector as a function of emission angle $\theta$ and kinetic energy $E_{\text{kin}}$ of the photoelectrons. The emission angle $\theta$ is related to the in-plane momentum of the electrons inside the solid via $k_{||}=\frac{\sqrt{2m_e}}{\hbar}\sin\theta\sqrt{E_{\text{kin}}}$, where $m_e$ is the electron mass. The binding energy $E_B$ of the electrons inside the solid can be obtained from $E_B=\hbar\omega-E_{\text{kin}}-\phi$, where $\hbar\omega$ is the photon energy and $\phi$ is the work function of the analyzer, that --- in the absence of space charge effects --- determines the kinetic energy where the Fermi level appears in the measurement. The energy resolution of the tr-ARPES measurements was determined from Fermi-Dirac fits of the carrier distribution inside the Dirac cone at negative pump-probe delay. The fitting function consisted of the Fermi-Dirac distribution convolved with a Gaussian to account for the finite energy resolution. These fits revealed that the kinetic energy at which the Fermi level appeared in the measurement (Fig.\,\ref{SFig2}a) as well as the energy resolution (Fig.\,\ref{SFig2}b) depend on the applied pump fluence. The temporal resolution of the tr-ARPES measurements of 200\,fs was deduced from the full width at half maximum (FWHM) of the width of the derivative of the rising edge of the photocurrent integrated over the area marked by the red box in Fig.\,1f. The conversion of the measured photocurrent from $(E_{\text{kin}}, \theta)$ to $(E_B, k_{||})$ relies on an accurate value for the kinetic energy of the Fermi level that is commonly used as energy reference in ARPES data. To solve this issue, we extrapolated the kinetic energy of the Fermi level in Fig.\,\ref{SFig2}a to zero pump fluence, yielding $E_{F,0}=17.48$\,eV.

\subsection{tr-ARPES data analysis}

\subsubsection{treatment of raw data}

\noindent The qualitiy of the tr-ARPES raw data in the present study was impaired by three main factors: (1) a featureless background due to dark counts on the detector as well as photoelectrons emitted from the pump rather than the probe pulse, (2) fluctuating ARPES counts due to shot-to-shot intensity variations of the XUV pulses, and (3) variations of the relative intensity of WS$_2$ and graphene bands due to sample inhomogeneity. These problems were solved as follows: 

\noindent (1) We measured the photocurrent with pump but without probe pulse and subtracted this background from the tr-ARPES snapshots. 

\begin{figure}
	\center
		\includegraphics[width = 1\columnwidth]{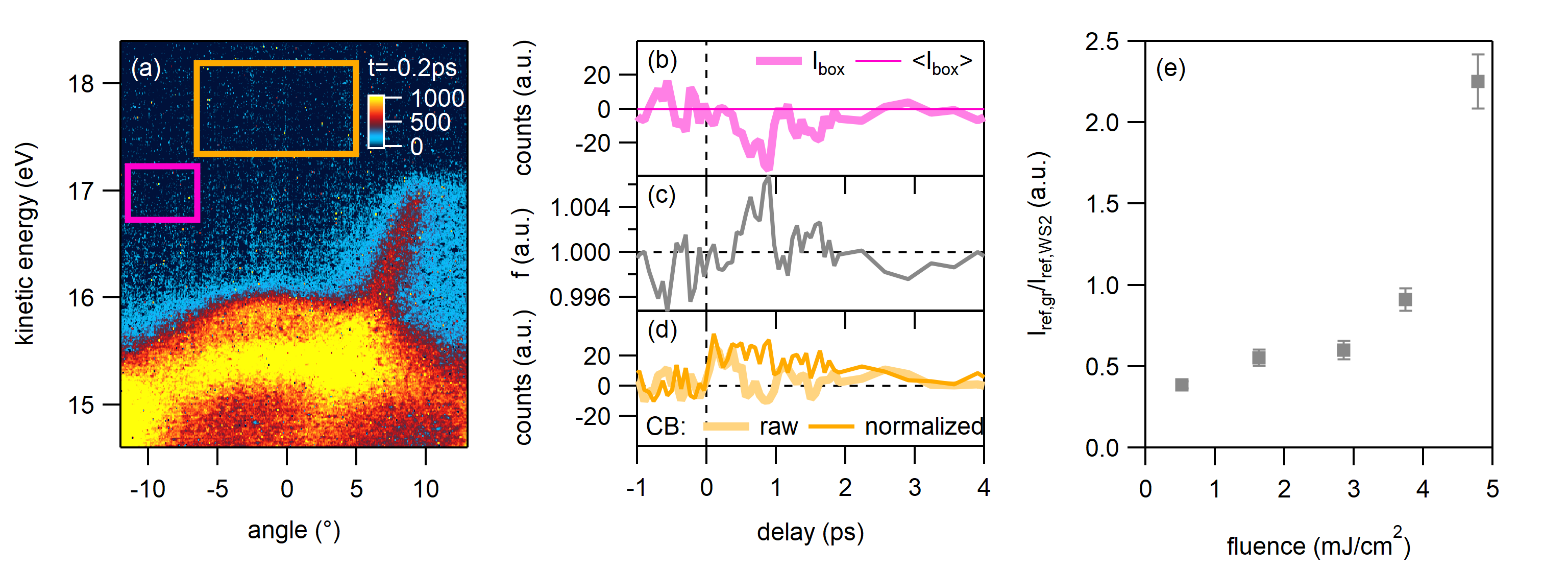}
  \caption{\textbf{Treatment of raw data:} (a) Tr-ARPES snapshot at negative pump-probe delay taken with a pump fluence of 1.63\,mJ/cm$^2$. The pink box is used to extract $I_{box}(t)$ in panel (b). The orange box captures the carriers at the bottom of the conduction band shown in panel (d). (b) Photocurrent $I_{box}(t)$ integrated over the area marked by the pink box in (a) together with its average for negative pump-probe delays $<I_{box}>$. (c) Normalization factor $f=<I_{box}>/I_{box}(t)$ as a function of pump-probe delay. (d) Effect of the raw data correction on the time dependence of the photocurrent integrated over the area marked by the orange box in panel (a). (e) Ratio of the reference intensities I$_{ref,gr}$ and I$_{ref,WS2}$ as defined in Fig.\,\ref{SFig4}c and a, respectively, for the datasets obtained with different pump fluences.}
  \label{SFig3}
\end{figure}

\noindent (2) We integrated the counts over the area marked by the pink box in Fig.\,\ref{SFig3}a where the photocurrent is expected to be zero after background subtraction. This yields a photocurrent $I_{box}(t)$ that fluctuates with time $t$ (Fig.\,\ref{SFig3}b). Next, we averaged $I_{box}(t)$ at negative delays yielding $<I_{box}>$. We then multiplied each tr-ARPES snapshot with a factor $f=<I_{box}>/I_{box}(t)$ (Fig.\,\ref{SFig3}c). The effect of this procedure on the transient photocurrent at the bottom of the conduction band obtained by integrating the tr-ARPES data over the area marked by the orange box in Fig.\,\ref{SFig3}a is illustrated in Fig.\,\ref{SFig3}d. 

\begin{figure}
	\center
		\includegraphics[width = 1\columnwidth]{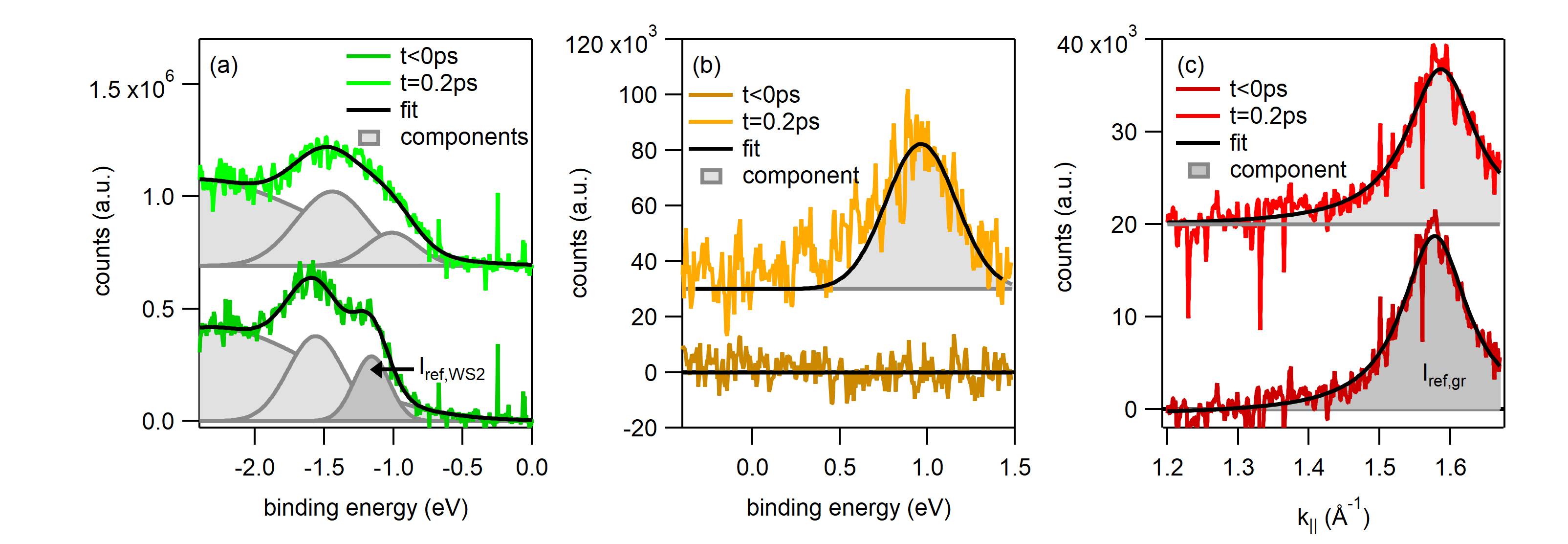}
  \caption{\textbf{Determination of transient band positions:} (a) EDCs extracted along the dashed green line in Fig.\,1a at $t<0$\,fs (light green) and at $t=200$\,fs (dark green) together with Gaussian fit (black) in the WS$_2$ VB region. The shaded areas indicate the three individual Gaussians used to fit the data. The area of the dark-shaded Gaussian I$_{ref,WS2}$ was used to turn the measured photocurrent independent of the WS$_2$ coverage. (b) EDCs extracted along the dashed green line in Fig.\,1a from the differential tr-ARPES data shown in the lower panel of Fig.\,1 in the WS$_2$ CB region for $t<0$\,fs (dark orange) and $t=200$\,fs (light orange) together with Gaussian fit (black). (c) MDCs extracted along the dashed red line in Fig.\,1a for $t<0$\,fs (dark red) and $t=200$\,fs (light red) together with Lorentzian fit. The area of the dark-shaded Lorentzian I$_{ref,gr}$ was used to turn the measured photocurrent independent of the WS$_2$ coverage. The data was obtained for a pump fluence of 2.85\,mJ/cm$^2$.}
  \label{SFig4}
\end{figure}

\noindent (3) As the WS$_2$ coverage in our WS$_2$/graphene heterostructure is not uniform, the relative intensity of the WS$_2$ and the graphene bands varies from spot to spot and, hence, between different tr-ARPES runs. In order to turn our results independent of the precise value of the WS$_2$ coverage we divided the pump-probe traces shown in Fig.\,2a and c by the integrated intensity of the upper WS$_2$ valence band I$_{ref, WS_2}$ from Fig.\,\ref{SFig4}a and the pump-probe trace shown in Fig.\,2b by the integrated intensity of the Dirac cone I$_{ref, gr}$ from Fig.\,\ref{SFig4}c. The ratio $I_{ref, gr}/I_{ref, WS_2}$ for the different tr-ARPES runs with different pump fluences is shown in Fig.\,\ref{SFig3}c. The measurement with the highest pump fluence was done on a sample area where the WS$_2$ coverage was particularly low. This reduced the absolute number of photoexcited electron-hole pairs and, hence, the absolute number of holes that were transferred to the graphene layer as well as the measured shift of the Dirac cone. For this reason, the high fluence data point in Fig.\,\ref{SFig9}c and d was set in brackets and not included for the determination of the guide to the eye.

\subsubsection{determination of transient band positions}

\noindent In Fig.\,\ref{SFig4} we show examples for the fits used to determine the transient peak positions in Fig.\,2. The EDCs in Fig.\,\ref{SFig4}a were obtained by integrating the tr-ARPES data in the first row of Fig.\,1 over a momentum interval of 0.05\,\AA$^{-1}$ centered at the green dashed line in Fig.\,1a. A total of three Gaussians (one for the upper and lower WS$_2$ VB, respectively, and one for the background) was used to fit the spectra. While the lower and upper VB are nicely resolved at negative pump-probe delays, the broadening of the spectrum after photoexcitation makes a correct determination of the transient peak positions difficult. This difficulty could be avoided by constraining the parameters of the fit as follows. (1) No peak was allowed to gain spectral weight with respect to negative pump-probe delay. (2) Only the upper valence band (VB$_A$) was allowed to loose spectral weight as a result of photoexcitation. (3) No peak was allowed to narrow with respect to negative pump-probe delay. (4) The separation of the upper and lower WS$_2$ was fixed at 430\,meV. (5) The position of the third Gaussian for the background was fixed at $E=-2.5$\,eV. The EDCs in Fig.\,\ref{SFig4}b were obtained by integrating the differential tr-ARPES data in the second row of Fig.\,1 over a momentum interval of 0.1\,\AA$^{-1}$ centered at the green dashed line in Fig.\,1a. These spectra were fitted with a single Gaussian for all pump-probe delays where the CB was found to be occupied. The integration range for the MDCs in Fig.\,\ref{SFig4}c was $\pm0.06$\,eV around $E=-0.5$\,eV. The MDCs were fitted with a single Lorentzian peak.

\subsubsection{Fermi-Dirac fits of electron distribution inside Dirac cone}

\begin{figure}
	\center
		\includegraphics[width = 0.5\columnwidth]{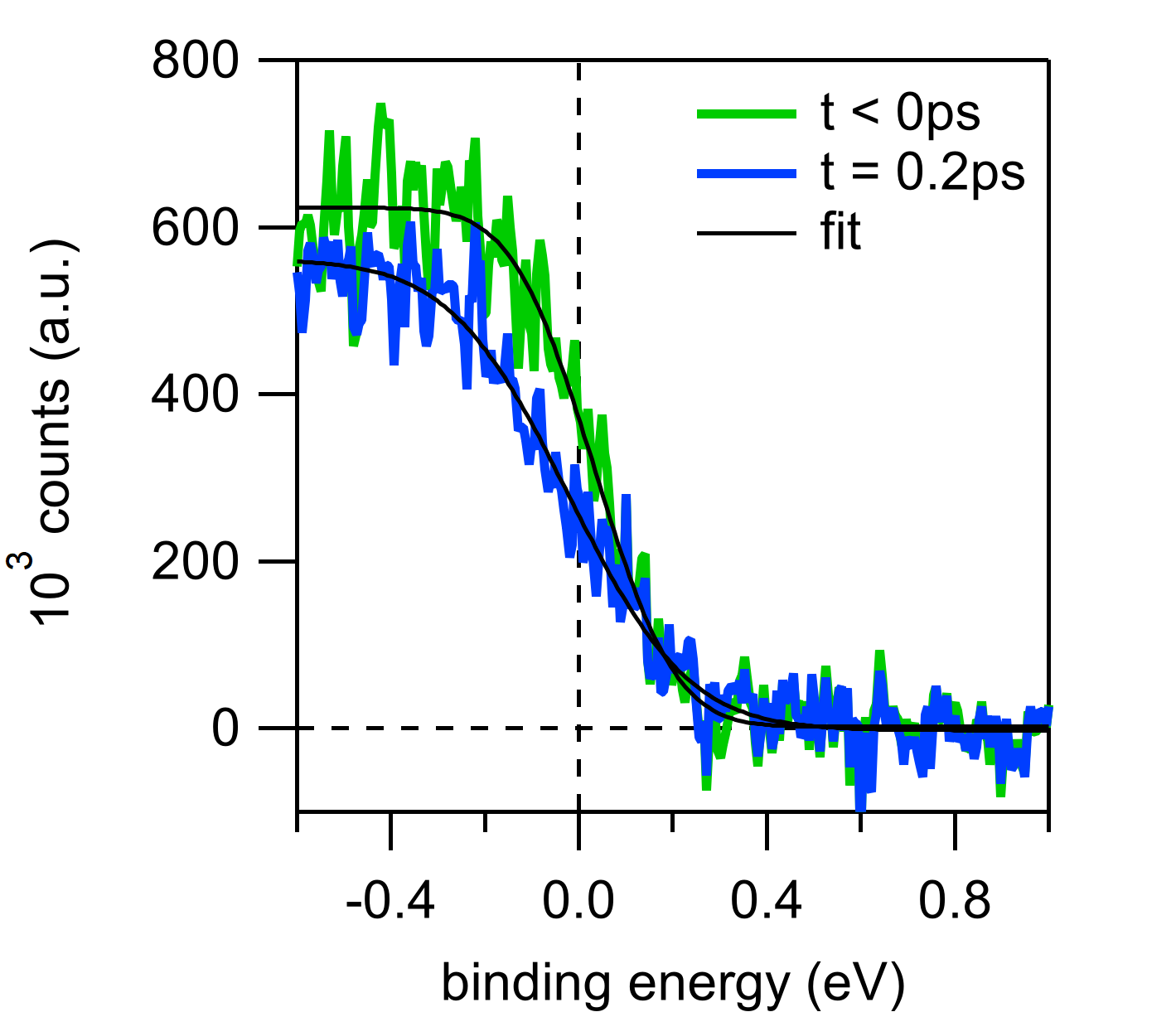}
  \caption{\textbf{Fermi-Dirac fits:} EDCs obtained by integrating the counts over the momentum width of the yellow box in Fig.\,1a at $t<0$\,fs (color) and $t=200$\,fs (color) together with Fermi-Dirac fits (black).}
  \label{SFig5}
\end{figure}

\noindent In Fig.\,\ref{SFig5} we provide an example for Fermi-Dirac fits of the transient electron distribution inside Dirac cone used to determine the electronic temperature and the chemical potential in Fig.\,\ref{SFig8}. The fitting function consisted of a Fermi-Dirac distribution convolved with a Gaussian to account for the finite energy resolution. From these fits the number holes transferred from the WS$_2$ into the graphene layer was determined as follows. The transient shift of the chemical potential referenced with respect to the graphene Dirac point $\mu_{e (ED)}$ was calculated by subtracting the band shift in Fig.Fig.\,2c from the chemical potential in Fig.\,\ref{SFig8}b that is referenced with respect to the vacuum level. From $\mu_{e (ED)}(t)$ and $T_e(t)$ (Fig.\,\ref{SFig8}a) we can then directly calculate the change of the total number of electrons in the graphene layer via

\begin{equation*}
\Delta n_e(t)=\int_{-\infty}^{\infty}{dE\,\rho(E)\left[f_{FD}(E,\mu(t),T(t))-f_{FD}(E,\mu_{0},T_{0})  \right]}
\end{equation*}

\noindent where $\rho(E)=\frac{2A_c}{\pi}\frac{|E-ED|}{\hbar^2 v_F^2}$ is the density of states with the unit cell are $A_c=\frac{3\sqrt{3}a^2}{2}$ and the lattice constant $a=1.42$\,\AA. The equilibrium chemical potential is $\mu_0=-0.3$\,eV. The transient chemical potential is given by $\mu(t)=\mu_0+\Delta \mu_{e (ED)}(t)$. The equilibrium temperature is $T_0=T(t<0\,\text{ps})=300$\,K. The number of transferred holes shown in Fig.\,\ref{SFig8}c is then given by $\Delta n_h(t)=-\Delta n_e(t)$.

\subsubsection{additional tr-ARPES data for other pump fluences}

\begin{figure}
	\center
		\includegraphics[width = 1\columnwidth]{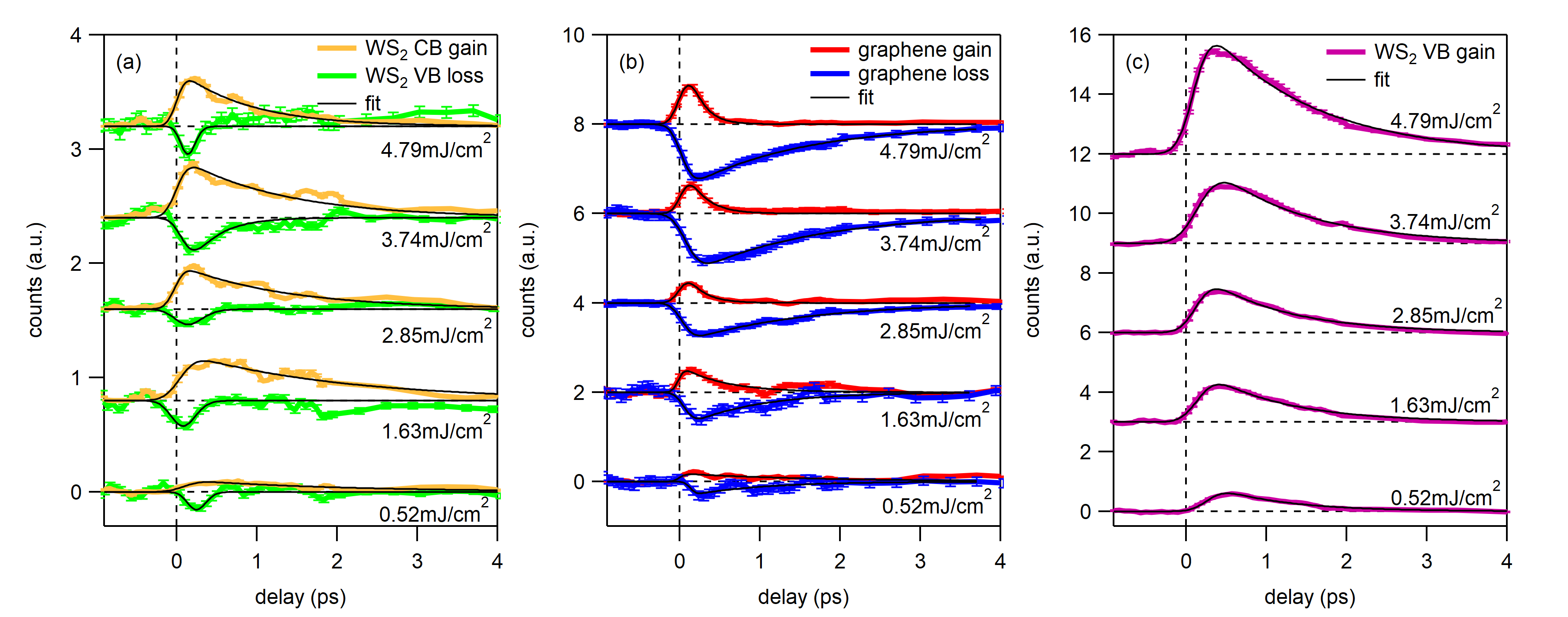}
  \caption{\textbf{Fluence dependence of the population dynamics:} (a) Gain in the CB (orange) and loss in the VB (green) of WS$_2$. (b) Graphene dynamics: Gain above the Fermi level (red) and loss below the Fermi level (blue). (c) Gain above the equilibrium position of the upper WS$_2$ VB. Thin black lines are single exponential fits to the data.}
  \label{SFig6}
\end{figure}

\begin{figure}
	\center
		\includegraphics[width = 1\columnwidth]{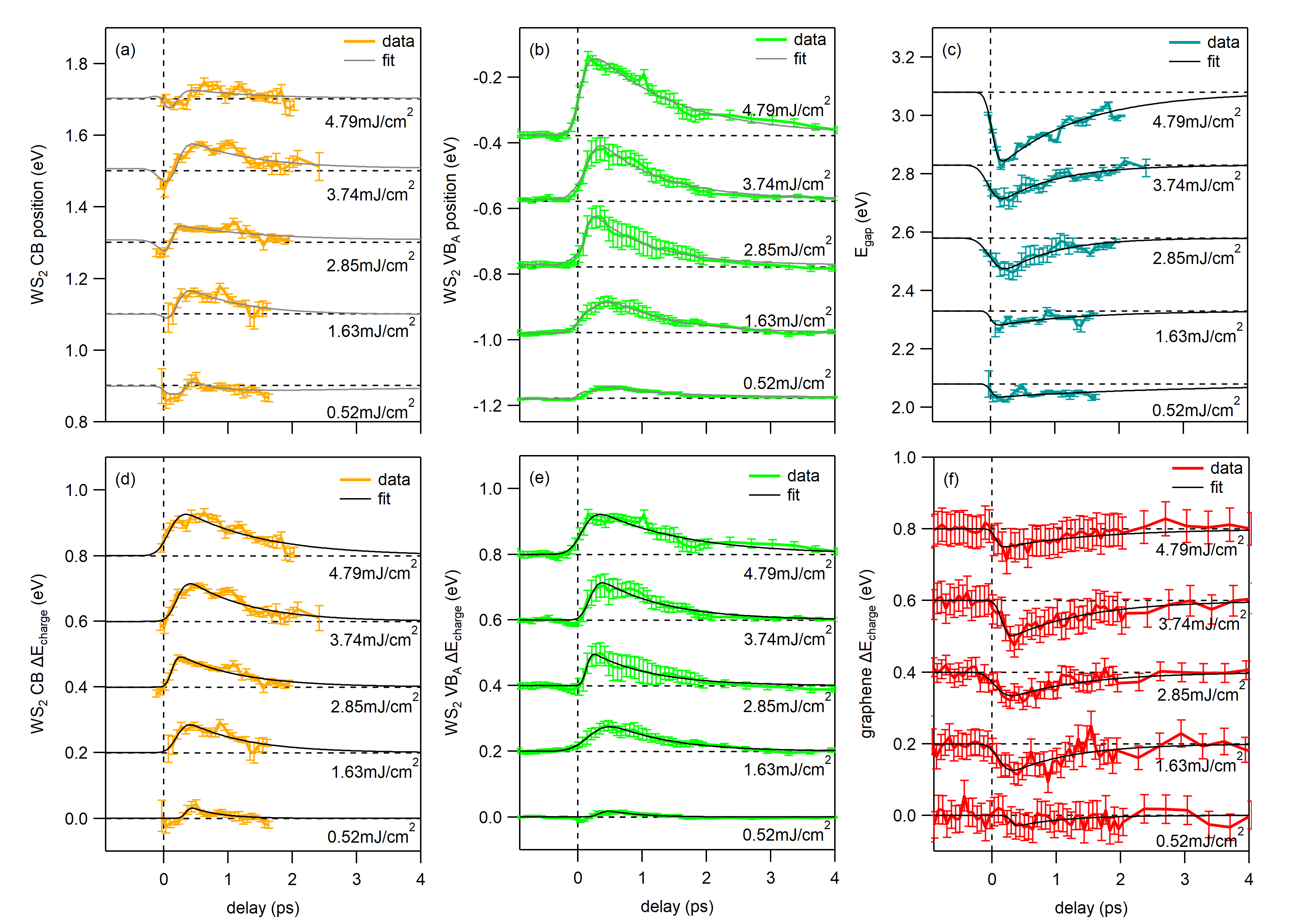}
  \caption{\textbf{Fluence dependence of the transient band positions:} (a) Position of the WS$_2$ CB. (b) Position of the WS$_2$ VB. (c) Transient band gap of  WS$_2$. (d) Shift of the WS$_2$ CB due to charging of the WS$_2$ layer. (e) Shift of the upper WS$_2$ VB due to charging of the WS$_2$ layer. (f) Shift of the Dirac cone.}
  \label{SFig7}
\end{figure}

\begin{figure}
	\center
		\includegraphics[width = 1\columnwidth]{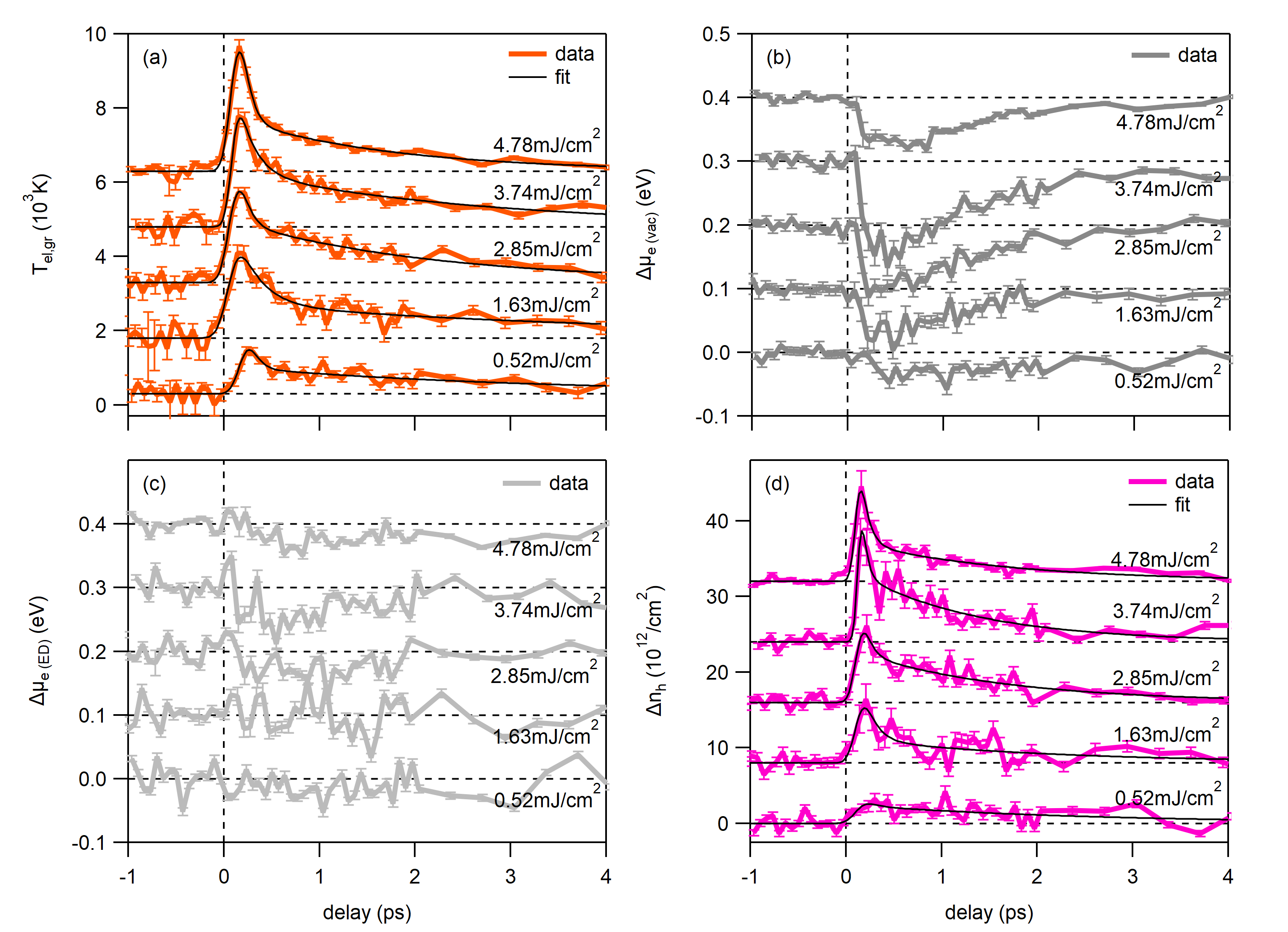}
  \caption{\textbf{Fluence dependence of hot carrier dynamics in graphene:} (a) Electronic temperature, (b) chemical potential referenced with respect to the vacuum level, (c) chemical potential referenced with respect to the Dirac point, and (d) number of holes transferred from WS$_2$ to graphene as a function of pump-probe delay for different pump fluences.}
  \label{SFig8}
\end{figure}

\begin{figure*}
	\center
		\includegraphics[width = 1\columnwidth]{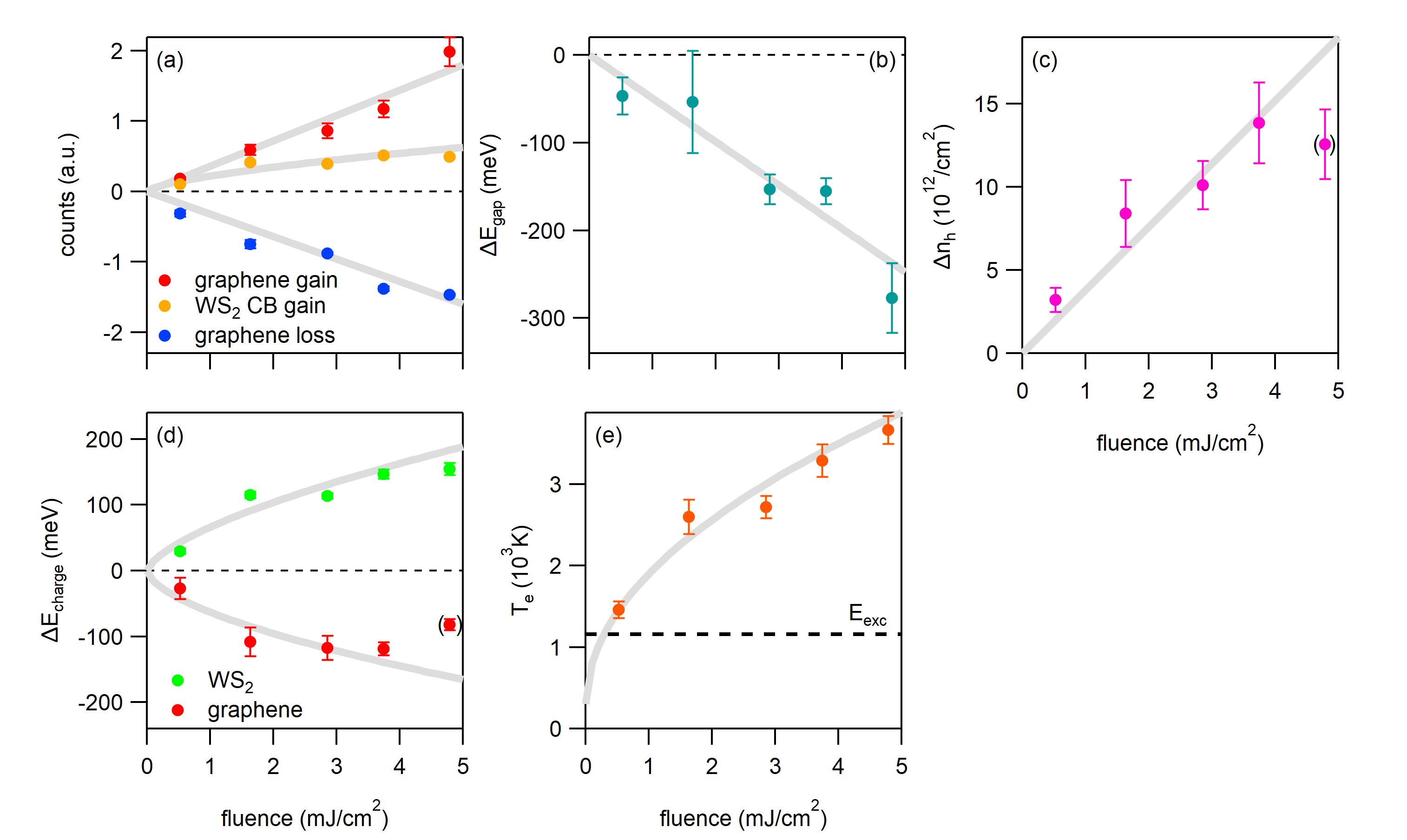}
  \caption{\textbf{Fluence dependence of different parameters at the peak of the pump-probe signal ($\bm{t=0.2}$\,ps):} (a) Pump-induced change of photocurrent inside colored boxes from Fig.\,1f. (b) Change of the WS$_2$ band gap from Fig.\,2a. (c) Number of holes transferred into the graphene layer from Fig.\,\ref{SFig8}c. (d) Charging-induced WS$_2$ and graphene band shifts from Figs.\,Fig.\,2e and f. (e) Peak electronic temperature of Dirac carriers from Fig.\,\ref{SFig8}a. Light gray lines are guides to the eye. The datapoints in brackets were obtained on areas of the sample with lower WS$_2$ coverage.}
  \label{SFig9}
\end{figure*}

\begin{figure}
	\center
		\includegraphics[width = 0.4\columnwidth]{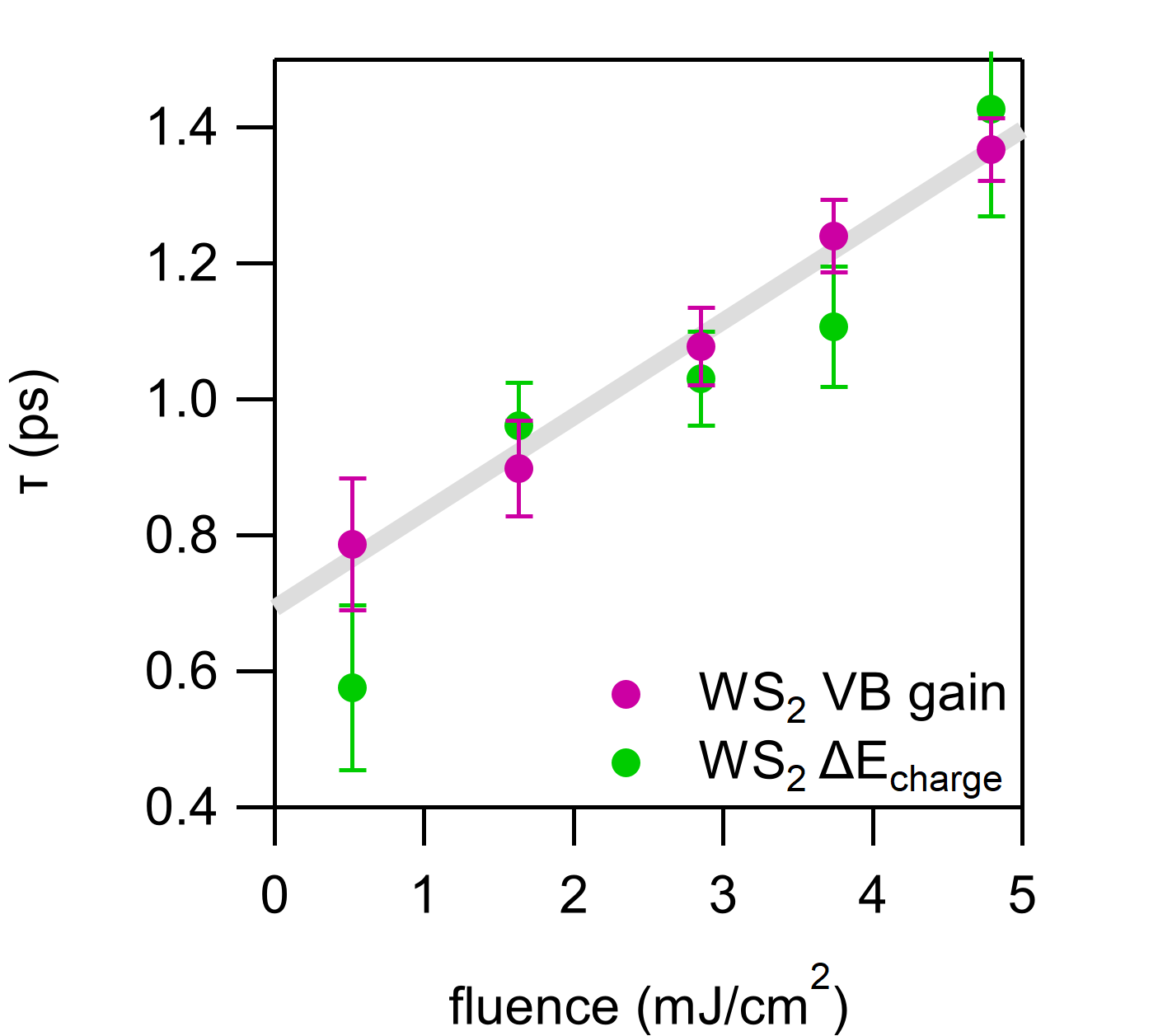}
  \caption{\textbf{Direct comparison} of fluence dependence of gain above equilibrium position of upper WS$_2$ VB (purple) and WS$_2$ charging shift (green).}
  \label{SFig10}
\end{figure}

\noindent The fluence dependent data corresponding to Fig.\,2 is shown in Figs.\,\ref{SFig6}-\ref{SFig7}. Figure\,\ref{SFig8} shows the fluence dependence of the hot carrier dynamics in graphene. The fluence dependence of various parameters at the peak of the pump-probe signal is shown in Fig.\,\ref{SFig9}. A comparison of the fluence dependence of the gain above the equilibrium position of the upper WS$_2$ VB and the charging-induced shift of the VB is shown in Fig.\,\ref{SFig10}.

\subsubsection{fit function for exponential decay}

\noindent The fitting function is given by

\begin{equation}
f(t)=\frac{a}{2} \left(1+\erf\left(\frac{(t-t_0)\tau-\sigma^2}{\sigma\,\tau}\right)\right)\exp\left(\frac{\sigma^2-2(t-t_0)\tau}{2\tau^2}\right)
\label{equ_fit}
\end{equation}

\noindent where $a$ is the amplitude of the pump-probe signal, $\sigma$ is related to the full width at half maximum (FWHM) of the derivative of the rising edge via $\text{FWHM}=2\sqrt{2\ln2}\,\sigma$, $t_0$ is the middle of the rising edge, $\erf$ is the error function, and $\tau$ is the exponential lifetime. This fitting function is obtained by convolving the product of a step function and an exponential decay with a Gaussian to account for the finite rise time of the signal. In Fig\,\ref{SFig10} we show that the gain above the equilibrium position of the upper WS$_2$ VB from Fig.\,2c and the WS$_2$ charging shift from Fig.\,2b exhibit the same fluence dependence. Hence, the lifetime of the gain above the equilibrium position of the upper WS$_2$ VB is a measure for the lifetime of the charge separated state.

\subsubsection{evidence for in-gap defect states}

\begin{figure}
	\center
		\includegraphics[width = 0.4\columnwidth]{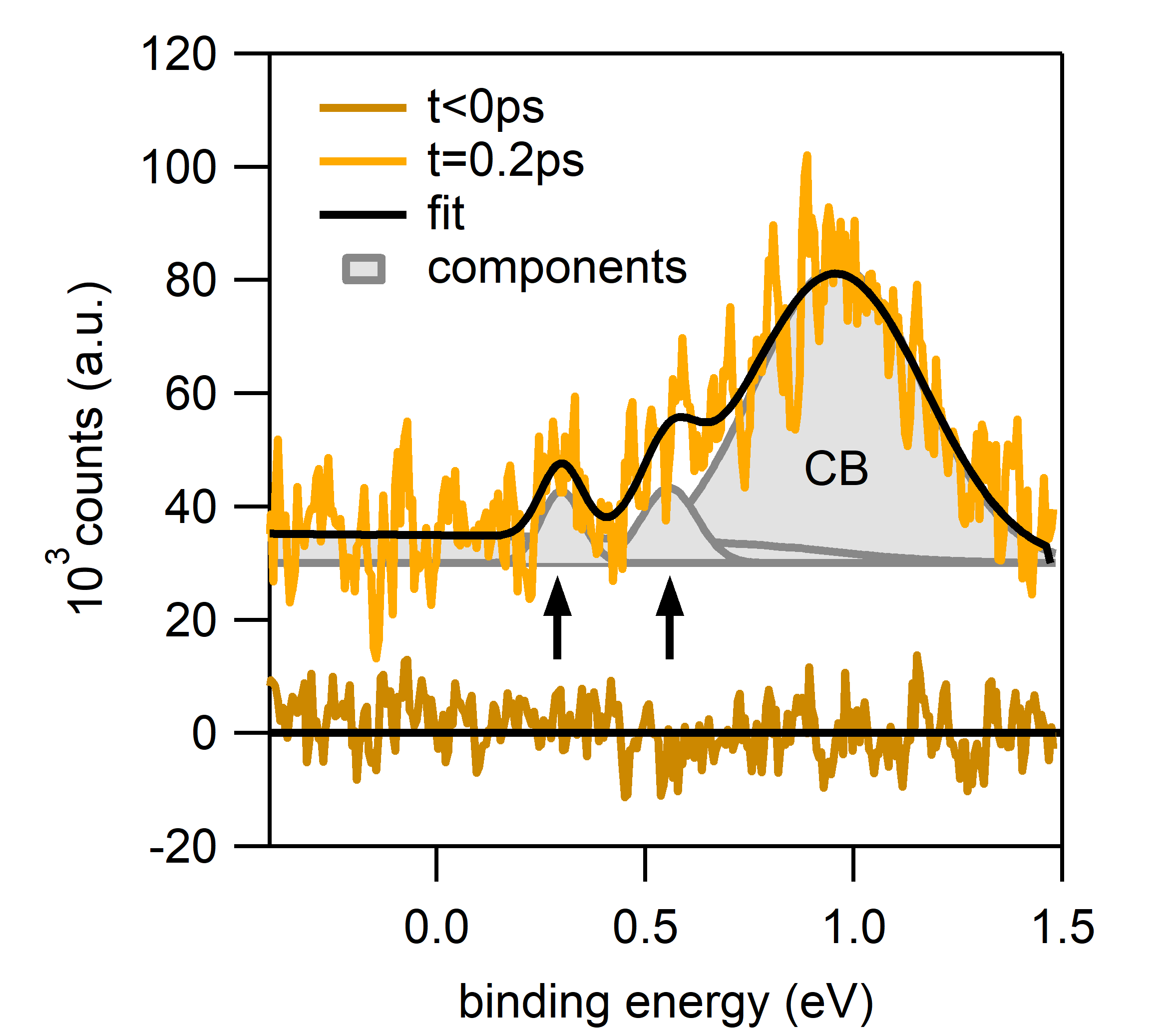}
  \caption{\textbf{Evidence for in-gap defect states.} EDCs through the K-point of WS$_2$ in the conduction band region at negative delay and at the peak of the pump-probe signal at $t=0.2$\,ps. Gray lines indicate the Gaussian peaks used to fit the spectrum at $t=0.2$\,ps. The two Gaussians marked with black arrows are attributed to in-gap defect states due to Selenium vacancies.}
  \label{SFig11}
\end{figure}

\noindent In Fig.\,\ref{SFig11} we show an energy distribution curve through the K-point of WS$_2$ in the energy region of the conduction band. At negative delay the WS$_2$ conduction band is unoccupied. At $t=0.2$\,ps we observe a pronounced Gaussian peak at $E=0.96$\,eV for the transiently populated WS$_2$ conduction band, as well as two other smaller Gaussians at $E=0.29$\,eV and $E=0.54$\,eV that we attribute to Selenium vacancies in good agreement with scanning tunneling spectroscopy data in Ref. \cite{SchulerPRL2019}.

\noindent


\section{III Comparison with literature}

\noindent In this section we compare our tr-ARPES data to other pump-fluence dependent experimental studies on similar vdW heterostructures \cite{UlstrupACSNano2016,SongOptik2018} and discuss whether the models proposed in literature \cite{LongNanoLett2016,WangNatCommun2016,ZhengNanoLett2017} might also apply in our case. 

\noindent Reference \cite{UlstrupACSNano2016} investigated an epitaxial MoS$_2$/graphene heterostructure on SiC(0001) with tr-ARPES. They observed a considerable band gap renormalization that increased with increasing hole density with a maximum value of $-450$\,meV at $n_h=1.5\times10^{-12}$\,cm$^{-1}$ with pump fluences in the mJ/cm$^{2}$ regime. These values are of the same order of magnitude as our data presented in Fig.\,\ref{SFig9}b. Unlike the present study, Ref. \cite{UlstrupACSNano2016} did not observe any indication for charge separation. We speculate that this might be related to the azimuthal alignment between the TMD and the graphene lattice which was 30$^{\circ}$ in \cite{UlstrupACSNano2016} and which is 0$^{\circ}$ or 60$^{\circ}$ in the present study.

\noindent Reference \cite{SongOptik2018} performed transient absorption measurements on CVD-grown WS$_2$/graphene heterostructues on SiO$_2$ but observed no systematic pump fluence dependence of the relaxation times with pump fluences in the $\mu$J/cm$^2$ regime. The authors proposed a model where the electric field across the interface that builds up due to charge separation caused by ultrafast hole transfer from WS$_2$ to graphene increases the transfer rate for the electrons which can be reconciled with our observation in Fig.\,3b. However, this model cannot explain our observation that the hole transfer rate also increases with increasing fluence (see Fig.\,3a).

\noindent References \cite{LongNanoLett2016,WangNatCommun2016,ZhengNanoLett2017} proposed a coherent phonon-driven charge transfer mechanism for ultrafast charge separation in vdW heterostructures with type II band alignment. The frequency of the associated coherent oscillations, however, is too high to be resolved given the limited temporal resolution of 200\,fs in the present study.

\section{IV Microscopic model of charge transfer}

\noindent In the second quantization formalism, the Hamilton operator describing electron tunneling from one layer ($l$) to another ($\bar{l}$) is written as
\begin{equation}
 H_T = \sum_{l \lambda \bm{k q}} T^{\lambda l \bar{l}}_{\bm{k q}} a^{\dagger}_{\lambda, \bar{l}, \bm{k} + \bm{q}} a^{\phantom{\dagger}}_{\lambda, l, \bm{k}},
\end{equation}
\noindent where $\lambda=v,c$ is the valence/conduction band, $\bm{k}$ the momentum of the initial state and $\bm{q}$ the momentum transfer of the process. The tunneling matrix element reads $T^{\lambda l \bar{l}}_{\bm{k q}} = \langle\bar{l},\lambda,\bm{k}+\bm{q} | V_T | l,\lambda,\bm{k}\rangle$ with the tunneling potential $V_T = V_l + V_{\bar{l}}$ being the sum of the potential of each layer. In order to find an expression for $T^{\lambda, \text{WS}_2 \rightarrow \text{g}}_{\bm{k q}}$, we use a tight-binding approach to describe the electronic wavefunctions in graphene and WS$_2$ \cite{SeligPRB2019}
\begin{equation}
    \psi^{\text{g}}_{\lambda \bm{k}}(\bm{r}) = \frac{1}{\sqrt{N_{\text{g}}}} \sum_{j=A,B} c^j_{\lambda \bm{k}} \sum_{\bm{R}_j} \mathrm{e}^{i \bm{k}\cdot\bm{R}_j} \phi^{\text{g}}_{\lambda}(\bm{r}-\bm{R}_j),
\end{equation}
\begin{equation}
    \psi^{\text{WS}_2}_{\lambda \bm{k}}(\bm{r}) = \frac{1}{\sqrt{N_{\text{WS}_2}}} \sum_{\bm{R}_{\text{W}}} \mathrm{e}^{i \bm{k}\cdot\bm{R}_{\text{W}}} \phi^{\text{WS}_2}_{\lambda}(\bm{r}-\bm{R}_{\text{W}}).
\end{equation}
\noindent Here $N_l$ is the number of unit cells in layer $l$, $c^j_{\lambda \bm{k}}$ are graphene's tight-binding coefficients, $\bm{R}_j$ are the atom positions, and $\phi^l_{\lambda}(\bm{r})$ is the linear combination of atomic orbitals contributing to the band $\lambda$. Graphene's tight binding coefficients around the $\bm{K}$ point read $c^A_{\lambda \bm{k}} = \frac{1}{\sqrt{2}}$ and $c^B_{\lambda \bm{k}} = \frac{1}{\sqrt{2}} \sigma_{\lambda} \mathrm{e}^{i \theta_{\bm{k}-\bm{K}_{\text{g}}}}$, where $\theta_{\bm{k}-\bm{K}_{\text{g}}}$ is the angle of $\bm{k}$ with respect to graphene's $\bm{K}$ point. While for graphene the wavefunction has contributions from the $A$ and $B$ lattice points, in semiconducting TMDs the main orbital contributions come from the metal atoms\,\cite{Kormanyos2DMater2015,WangPhysRevB2017}. Introducing the respective electronic wavefunctions, the tunneling matrix element reads
\begin{align}
    T^{\lambda, \text{WS}_2 \rightarrow \text{g}}_{\bm{k q}} =& \int d^3\bm{r} \left[ \psi^{\text{g}}_{\lambda \bm{k}+\bm{q}}(\bm{r}) \right]^* V_T(\bm{r}) \psi^{\text{WS}_2}_{\lambda \bm{k}}(\bm{r}) \nonumber \\
    =& \frac{1}{\sqrt{N_g N_{\text{WS}_2}}} \sum_{j \bm{R}_j \bm{R}_{\text{W}}} \left( c^j_{\lambda \bm{k}+\bm{q}} \right)^* \mathrm{e}^{- i (\bm{k}+\bm{q})\cdot\bm{R}_j} \mathrm{e}^{i \bm{k}\cdot\bm{R}_{\text{W}}} \nonumber \\
    & \times \int d^3\bm{r} \left[ \phi^{\text{g}}_{\lambda}(\bm{r}-\bm{R}_j) \right]^* V_T(\bm{r}) \phi^{\text{WS}_2}_{\lambda}(\bm{r}-\bm{R}_{\text{W}})
\end{align}

\noindent Using $V_T = V_l + V_{\bar{l}}$ and the periodicity of $V_l$ in the lattice of $l$ and writing the orbital wave functions in terms of their in-plane Fourier transform, $\phi(\bm{r}) = \frac{1}{A} \sum_{\bm{p}} \mathrm{e}^{i \bm{p}\cdot\bm{r}_{\parallel}} \phi_{\bm{p}}(z)$, we can define a tunneling parameter $h_{\lambda \bm{p}}$ that contains the overlap between the orbital wave functions and the tunneling potential. The tunneling matrix element now reads
\begin{equation}
    T^{\lambda, \text{WS}_2 \rightarrow \text{g}}_{\bm{k q}} = \frac{1}{N_g N_{\text{WS}_2}} \sum_{j \bm{R}_j \bm{R}_{\text{W}} \bm{p}} \left( c^j_{\lambda \bm{k}+\bm{q}} \right)^* \mathrm{e}^{- i (\bm{k}+\bm{q}-\bm{p})\cdot\bm{R}_j} \mathrm{e}^{i (\bm{k}-\bm{p})\cdot\bm{R}_{\text{W}}} h_{\lambda \bm{p}},
\end{equation}
\noindent with the tunneling parameter
\begin{equation}
    h_{\lambda\bm{p}} = t_{\lambda\bm{p}}^{\text{WS}_2,\text{g}} + \left( t_{\lambda\bm{p}}^{\text{g},\text{WS}_2} \right)^*, \qquad t_{\lambda\bm{p}}^{l\bar{l}} = \frac{1}{\sqrt{N_g N_{\text{WS}_2}}} \int d^3\bm{r} \mathrm{e}^{i \bm{p}\cdot\bm{r}_{\parallel}} \left[ \phi^{\bar{l}}_{\lambda\bm{p}}(z) \right]^* V_l(\bm{r}) \phi^{l}_{\lambda}(\bm{r})
\end{equation}
\noindent Now we can perform the summation over the lattice points of each layer using $\frac{1}{N_l}\sum_{\bm{R}_{lj}} \mathrm{e}^{i \bm{k}\cdot\bm{R}_{lj}} = \sum_{G_l} \mathrm{e}^{i \bm{G}_l\cdot\bm{\delta}_{jl}} \delta_{\bm{k},\bm{G}_l}$ and determine the allowed momentum exchange $\delta_{\bm{q},\bm{G}_{\text{g}}-\bm{G}_{\text{WS}_2}}$, where $\bm{G}_l$ is the reciprocal lattice vector of the layer $l$. We have introduced here the lattice offsets $\bm{\delta}_{jl}$, which we define as $\bm{\delta}_{A} = - \bm{\delta}_{\text{W}} = \frac{1}{2}\bm{R}_A^0$ and $\bm{\delta}_{A} = \frac{1}{2}\bm{R}_A^0 + \bm{\tau}_{AB} = \frac{1}{2}\bm{R}_B^0$, with $\bm{\tau}_{AB}$ being a vector from an $A$ lattice point to a nearest neighbor $B$. The tunneling matrix element now reads
\begin{equation}
    T^{\lambda, \text{WS}_2 \rightarrow \text{g}}_{\bm{k q}} = \sum_{j \bm{G}_{\text{g}} \bm{G}_{\text{WS}_2}} \left( c^j_{\lambda, \bm{k}+\bm{q}} \right)^* \mathrm{e}^{i \frac{1}{2} (\bm{G}_{\text{g}}\cdot\bm{R}_j^0+\bm{G}_{\text{WS}_2}\cdot\bm{R}_A^0)} h_{\lambda, \bm{k}+\bm{G}_{\text{WS}_2}} \delta_{\bm{q},\bm{G}_{\text{g}}-\bm{G}_{\text{WS}_2}}.
\end{equation}
\noindent In a practical scenario, the initial momentum $\bm{k}$ will lie in the vicinity of $\bm{K}_{\text{WS}_2}$. Assuming that the tunneling parameter decays quickly with momentum, the significant reciprocal lattice vectors will be those connecting two K points of the Brillouin zone, i.e. $\bm{K}_l+\bm{G}_l=C^n_3\bm{K}_l$. Hence we can reduce the allowed scattering transitions to $\delta_{\bm{q}, C_3^n \Delta K - \Delta K}$. Finally, the rotational symmetry of the orbitals yields $h_{\lambda,C_3^n K_{\text{WS}_2}} = h_{\lambda,K_{\text{WS}_2}} \mathrm{e}^{i \varphi_n}$, with $\varphi_{\lambda,n} = \frac{2\pi}{3} n (M_{\lambda}^{\text{WS}_2}-M_{\lambda}^{\text{g}})$, where $M_{\lambda}^l$ is the rotational quantum number. Defining the Moiré phase $\Phi_{j,n} = \frac{1}{2} (\bm{G}_{\text{g}}\cdot\bm{R}_j^0+\bm{G}_{\text{WS}_2}\cdot\bm{R}_A^0)$, the tunneling matrix element now reads
\begin{equation}
    T^{\lambda, \text{WS}_2 \rightarrow \text{g}}_{\bm{k q}} = \sum_{j, n=0}^2 \left( c^j_{\lambda, \bm{k}+\bm{q}} \right)^* \mathrm{e}^{i \Phi_{j,n}} \mathrm{e}^{i \varphi_{\lambda,n}} h_{\lambda, \bm{K}_{\text{WS}_2}} \delta_{\bm{q},C_3^n \Delta K - \Delta K}.
\end{equation}
\noindent Now we choose the offset $\bm{R}_A^0=0$, resulting in $\Phi_{A,n} = 0$, $\Phi_{B,n} = - \frac{2\pi}{3} n$. Introducing graphene's tight-binding coefficients, we obtain
\begin{equation}
    T^{\lambda, \text{WS}_2 \rightarrow \text{g}}_{\bm{k q}} = \sum_{n=0}^2 \left( 1 + \sigma_{\lambda} \mathrm{e}^{i\theta_{\bm{k}+\bm{q}-\bm{K}_{\text{g}}}} \mathrm{e}^{- i \frac{2\pi}{3} n} \right)\mathrm{e}^{i \varphi_{\lambda,n}} h_{\lambda, \bm{K}_{\text{WS}_2}} \delta_{\bm{q},C_3^n \Delta K - \Delta K}.
\end{equation}

\noindent Finally, we are interested in the squared absolute value of the tunneling matrix element. Note that if the condition $\bm{q} = C_3^n \Delta K - \Delta K$ is fulfilled, this will be the case only for one $n$. Hence the phase $\varphi_{\lambda,n}$ will in fact be a global phase and cancels out when computing the absolute value. Thus, we can write our final expression for the tunneling matrix element as
\begin{equation}
 | T^{\lambda, \text{WS}_2 \rightarrow \text{g}}_{\bm{k q}} |^2 = \sum_{n=0}^2 |h_{\lambda,\bm{K}_{\text{WS}_2}}|^2 \left[ 1 + \sigma_{\lambda} \cos \left( \theta_{\bm{k}-\bm{K}_\text{g}+\bm{q}} - \frac{2\pi}{3}n \right) \right] \delta_{\bm{q}, C_3^n \Delta K - \Delta K}.
\end{equation}
\noindent Here $\sigma_{\lambda}=\pm 1$ for the conduction ($+$) and the valence ($-$) bands, $\theta_{\bm{k}-\bm{K}_{\text{g}}+\bm{q}}$ is the angle of $\bm{k}+\bm{q}$ with respect to the closest graphene $\bm{K}_\text{g}$ point, $\Delta K = K_{\text{g}} - K_{\text{WS}_2}$ is the momentum difference between graphene's and WS$_2$'s $K$ points, $C_3^n$ is a $2\pi/3$ rotation operator, and $h_{\lambda}$ contains the overlap of the wavefunctions with the tunneling potential. From this expression we obtain the behaviour shown in Fig.\,5(a)-(b), i.e. the tunneling is efficient for holes but suppressed for electrons. This effect, together with the cosine dependence, is a manifestation of graphene's pseudospin.

\noindent In order to calculate tunneling rates, we insert the tunneling Hamilton operator in the Heisenberg's equation together with the carrier occupation $\rho^l_{\lambda \bm{k}}$ in the density matrix formalism \cite{MalicPRB2011}. Within a second-order Born-Markov approximation\,\cite{KiraProg2016}, we find the following Boltzmann-like equation:
\begin{equation}
 \dot{\rho}^l_{\lambda \bm{k}} = \frac{2 \pi}{\hbar} \sum_{\bm{q}} | T^{\lambda l \bar{l}}_{\bm{k q}} |^2 \left[ \rho^{\bar{l}}_{\lambda,\bm{k}+\bm{q}} (1 - \rho^l_{\lambda,\bm{k}}) - \rho^l_{\lambda,\bm{k}} (1 - \rho^{\bar{l}}_{\lambda,\bm{k}+\bm{q}}) \right] \delta(\varepsilon^l_{\lambda, \bm{k}} - \varepsilon^{\bar{l}}_{\lambda, \bm{k}+\bm{q}}).
\end{equation}
\noindent Since we do not observe significant electron occupations at the conduction band (cf. Fig.\,1), the carrier occupation in WS$_2$ can be well described by a Boltzmann distribution. Integrating over momentum leads to a rate equation for the carrier density. We thus find that the carrier density that tunnels from WS$_2$ to graphene follows
\begin{equation}
 \left. \dot{n}^{\text{WS}_2}_{\lambda} \right|_{\text{WS}_2 \rightarrow \text{g}} = - \tau_{\lambda}^{-1} n^{\text{WS}_2}_{\lambda},
\end{equation}
\noindent where
\begin{equation}
    \tau_{\lambda}^{-1} = \frac{2\pi}{\hbar A n^{\text{WS}_2}_{\lambda}} \sum_{\bm{k}\bm{q}} | T^{\lambda l \bar{l}}_{\bm{k q}} |^2 \rho^{\text{WS}_2}_{\lambda,\bm{k}} (1 - \rho^{\text{g}}_{\lambda,\bm{k}+\bm{q}}) \delta(\varepsilon^{\text{WS}_2}_{\lambda, \bm{k}} - \varepsilon^{\text{g}}_{\lambda, \bm{k}+\bm{q}})
\end{equation}

\end{document}